\documentclass[conference]{IEEEtran}
\usepackage{booktabs}
\usepackage{balance}
\usepackage{graphics}
\usepackage[T1]{fontenc}
\usepackage{multirow}
\usepackage{enumitem}
\usepackage{graphicx}
\usepackage{mdwlist}
\usepackage{tabularx}
\usepackage{color}
\usepackage{soul}
\usepackage{ifthen}
\usepackage{comment}
\usepackage{bm}

\newcommand{\etal}{et~al.~}
\newcommand{\ie}{i.e.,~}
\newcommand{\eg}{e.g.,~}

\begin{document}

\linespread{0.9}
\title{Visual Analytics of Anomalous User Behaviors:\\A Survey}
\author{\small{Yang Shi$^{1}$, Yuyin Liu$^{2}$, Hanghang Tong$^{3}$, Jingrui He$^{3}$, Gang Yan$^{1}$, Nan Cao$^{1}$}\\
 $^1$Tongji University, China\\
         $^2$Imperial College London, United Kingdom\\
         $^3$University of Illinois at Urbana-Champaign, United States
}

\maketitle

\begin{abstract}
The increasing accessibility of data provides substantial opportunities for understanding user behaviors. Unearthing anomalies in user behaviors is of particular importance as it helps signal harmful incidents such as network intrusions, terrorist activities, and financial frauds. 
Many visual analytics methods have been proposed to help understand user behavior-related data in various application domains. 
In this work, we survey the state of art in visual analytics of anomalous user behaviors and classify them into four categories including social interaction, travel, network communication, and transaction. We further examine the research works in each category in terms of data types, anomaly detection techniques, and visualization techniques, and interaction methods. Finally, we discuss findings and potential research directions. 
\end{abstract}

\begin{figure*}[!t]
    \centering
    \includegraphics[width=\textwidth]{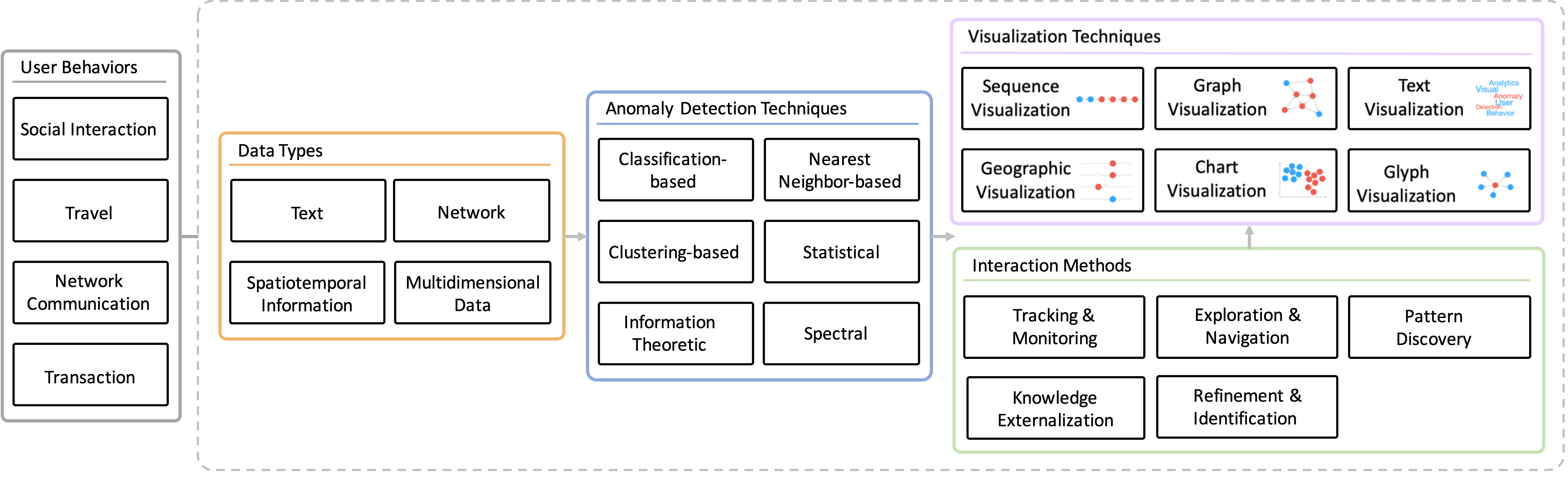}
    \caption{Taxonomy of this survey, addressing the data type, anomaly detection techniques~\cite{Chandola2009AnomalyDA}, visualization techniques, and interaction methods in the visual analysis of anomalous user behaviors.}
    \label{fig:overview} 
\end{figure*}

\section{Introduction}

The increasing accessibility of data collected from various sources provides potential opportunities for understanding user behaviors.
Identifying anomalies in user behaviors is of particular interest in many application domains such as cybersecurity, urban planning, and social media. For instance, detecting rumors and tracking their spreading patterns alert people to the risks of being influenced by misinformation, which is especially critical in political elections. 

Detecting anomalous user behaviors is a challenging task as the boundary between abnormal and normal data cannot be clearly defined. 
Even equipped with domain knowledge, analysts may find results of automatic machine learning approaches lack contextual information to support decision-making, \eg analysts are limited to exploring who did what when and where, why (5W's) and how. 
To address the issue, visualization integrates human knowledge into information processing tasks. It presents anomalous patterns intuitively to decision makers as well as involves a human-machine dialog as they interact with the data set.
Our work aims to summarize the-state-of-art in visual analytics of anomalous user behaviors, with the purpose of highlighting current research trends as well as future directions.

In this survey, we contribute a taxonomy of visual analytics of anomalous user behaviors. The overview of the analytical pipeline is summarized in Figure~\ref{fig:overview}. 

\begin{itemize}[leftmargin=*]
 
 \item We categorize four user behaviors, including social interaction, travel, network communication, and transaction based on the data collected from specific data sources. We extract four common data types from these four behaviors, including text, network, spatiotemporal information, and multidimensional data. 
 
 \item We review how research works use visualization techniques combined with interaction methods to analyze anomalous user behaviors. We extract six visualization techniques, including sequence visualization, graph visualization, text visualization, geographic visualization, chart visualization, and glyph visualization. We also summarize six interaction methods, including tracking \& monitoring, exploration \& navigation, pattern discovery, knowledge externalization, and refinement \& identification.

\end{itemize}

The remaining survey is organized as follows. First, we describe related surveys in Section~\ref{sec:related}. Then, we present the taxonomy, methodology, and taxonomy used in this survey in Section~\ref{sec:taxonomy}.
Section \ref{sec:social}, \ref{sec:travel}, \ref{sec:network}, and \ref{sec:trasaction} analyze the four user behaviors respectively using the taxonomies explained in Section \ref{sec:taxonomy}. Analysis of each behavior follows the general visual analytics pipeline. We start with identifying data types and anomaly detection techniques, visualization techniques and interaction methods are then discussed. 
Finally, we discuss findings and trends acquired from surveying papers in Section~\ref{sec:discussion} and conclude our work in Section~\ref{sec:conclusion}.

\section{Related Surveys}
\label{sec:related}
In this section, we discuss related surveys for visual anomalous user behaviors analysis. 
There are survey papers in the literature that focus on analyzing user behaviors.
Jin \etal~\cite{jin2013understanding} categorize user behaviors in online social network into four types including connectivity and interaction, traffic activity, mobile social behavior, and malicious behavior. 
Jiang \etal~\cite{jiang2016suspicious} classify anomalous behaviors when using web applications (\eg Hotmial, Facebook, Amazon) into four categories: traditional spam, fake reviews, social spam, and link farming. 
Surveys regarding visualization of user behaviors data explore application domains such as urban computing~\cite{zheng2016visual}, social media~\cite{chen2017social, wu2016survey}, financial domain \cite{ko2016survey}, and network security~\cite{shiravi2012survey, lavigne2014visual}. 
In the field of anomaly detection, Chandola \etal~introduce categories of anomaly detection (AD) techniques \cite{Chandola2009AnomalyDA}. 
\cite{Patcha2007AnOO} and \cite{Akoglu2014GraphBA} examine techniques used in intrusion detection systems and for detecting graph-based anomalies, respectively. 
Recent work of Chalapathy and Chawla~\cite{chalapathy2019deep} present a structured overview of research approaches in deep learning-based anomaly detection.
Our survey covers a wider range of application domains than existing surveys. 
To the best of our knowledge, it is the first survey that explores anomalous user behaviors from a perspective of visual analytics. 

\section{Terminology, Methodology, and Taxonomy}
\label{sec:taxonomy}
In this section, we first explain the terminology used in this survey and describe our methodology of selecting papers suitable for the topic of the survey. Next, we introduce the taxonomy of anomalous user behaviors regarding common data types, anomaly detection techniques, visualization, and interaction methods.  

\subsection{Terminology}
The survey aims to summarize visualization works that focus on anomalous user behaviors.
Here, \textit{user behaviors} can be derived directly and indirectly from user actions. For example, posting a tweet is a behavior directly related to user actions while a cyber-attack is conducted by nodes in networks but indirectly manipulated by the perpetrator. Investigation of user behavior focuses on tracking, collecting, and assessing patterns caused by users' as opposed to information of devices and events~\cite{litan2014market, rouse2017user}. 
Analyzing and identifying anomalous user behaviors uses anomaly detection techniques. 
According to Chandola \etal~\cite{Chandola2009AnomalyDA}, anomalies are \textit{``patterns in data that do not conform to a well-defined notion of normal behavior''}. As we collect research works from a diverse set of domains such as social media, finance, and cybersecurity, the scope of anomaly detection in our survey is broader than the scope identified in specific domains. For example, \eg Chen \etal~\cite{chen2017social} identify data outside normal ranges of attributes as anomalies in social media while in the field of cybersecurity, anomalies refer to malware, insider threats, and targeted attacks~\cite{litan2014market, rouse2017user}. 
In our work, \textit{anomalies} refer to frauds, spam, intrusion, sudden increases in the volume of data, and periodic patterns of users, etc. In short, as long as results detected express \textit{``interestingness of real-life relevance''}~\cite{Chandola2009AnomalyDA}, we claim that the visualization works are within the scope of \textit{anomaly detection}. 

\subsection{Methodology}
Our interested range of publications is constrained by three conditions: user behaviors, anomaly detection, and visual analytics/visualization. We started from a core set of relevant research works known to us in advance, and followed references from ``Related Work'' as well as papers that cite the previously identified papers. We also conducted a keyword search for papers published in visualization conferences or journals. Examples of keywords are ``anomaly, anomalous, outlier, abnormal, unusual'' and ``rare''. The research papers were checked to affirm that they are indeed associated with the concept of anomaly in~\cite{Chandola2009AnomalyDA}. The association with user behaviors was expected to be seen in Case Study section in publications. 
During the process of investigating research works, we found that the range of pertinent papers is relatively narrow.
To solve the potential shortage in the number of references, our survey range covered publications that incorporate anomaly detection as one of their visual analytic approaches in addition to those that solely address the issue of anomaly detection, \eg we include \cite{yeon2017predictive} in our collection through the authors' ultimate goal is predictive analysis of event evolution.

We also keep our exploration spectrum balanced in terms of application domains. We noticed the number of publications related to \textit{travel} and \textit{network communication} outnumber others. The outnumbering of travel probably results from the early history of visualizing spatiotemporal data (in 1869 Charles Minard produced a map to illustrate Napoleon's March to Moscow) and continuous study ever since. 
As for cybersecurity, the establishment of a conference for visualization of cybersecurity, \textit{IEEE Symposium on Visualization for Cyber Security (VizSec)}, encourages researchers to devote efforts in this field. 
As such, we allocated more time to searching for research works of other user behaviors comparatively. We are hoping to capture possibly interesting relationships across user behaviors by maintaining a broad scope of investigation.

\subsection{Taxonomy}
\begin{figure*}[!t]
    \centering
    \includegraphics[width=\textwidth]{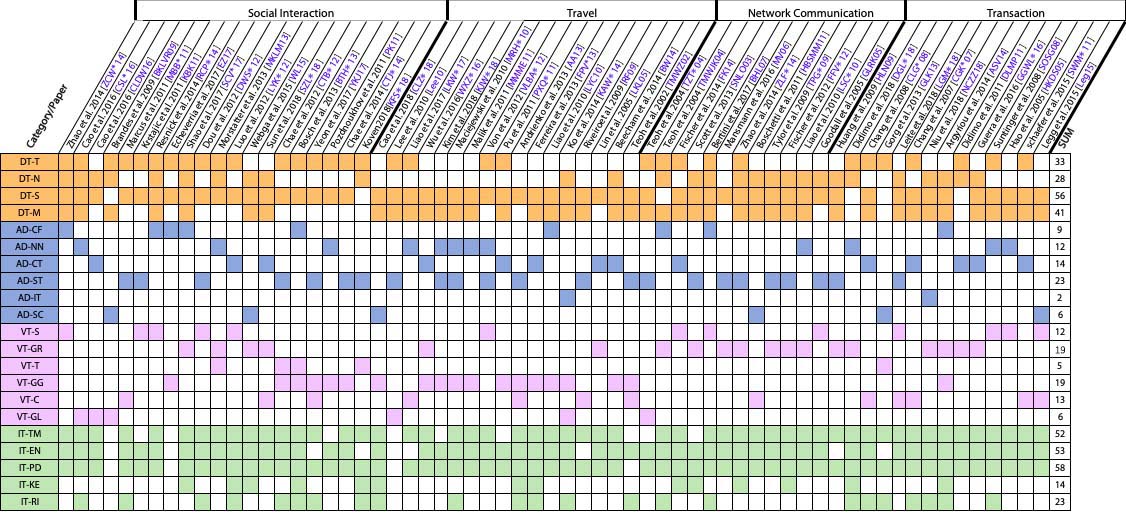}
    \caption{The selected papers regarding visualization and visual analytics of anomalous behaviors. DTs: text, network, spatiotemporal information, and multidimensional data. ADs: classification-based, nearest neighbor-based, clustering-based, statistical, information theoretic, and spectral anomaly detection techniques. VTs: sequence, graph, text, geographic, chart, and glyph visualizations. ITs: tracking \& monitoring, exploration \& navigation, knowledge externalization, pattern discovery, and refinement \& identification.}
    \label{tab:sum} 
\end{figure*}

Based on a literature review of more than 150 papers that relevant to visual analytics of anomalous user behaviors, we summarize four user behaviors including social interaction, travel, network communication, and transaction.
For each of the four user behaviors, we attempt to identify common data types, anomaly detection techniques, visualization, and interaction methods. 
The different categories are highlighted in the overall pipeline of visual analytics in Figure~\ref{fig:overview}. The selected papers are summarized in Table~\ref{tab:sum}, with color indicates each category.

\textbf{User Behaviors.}
User behaviors are seen in a variety of application domains. Based on the data collected form specific data sources, we classify user behaviors into four categories: social interaction, travel, network communication, and transaction. 
\textit{Social interaction} describes the communication of ideas and thoughts between people. Its data is collected from publicly accessible social platforms or private telecommunication platforms.
\textit{Travel} is the physical movement of users between places containing geographic information. Its data is collected from Global Positioning System (GPS), mobile phones and base stations, etc. 
\textit{Network communication} is sending and receiving information between machines via networks. Its data is collected from server logs. 
\textit{Transaction} refers to monetary flows in buying and selling, whose data is collected from system logs. 

We also categorize anomalous user behaviors into egocentric and collective behaviors. 
The categorization is inspired by the concepts of point and collective anomalies~\cite{Chandola2009AnomalyDA}. 
Note that our survey focuses on the investigation of anomalous user behaviors which constitutes a subset of anomalies. 
\textit{Egocentric} behavior refers to the user behavior that distinguishes itself from the rest of data in anomaly detection. \textit{Collective} behavior is a set of user behaviors that appear anomalous. When analyzing separate user behaviors categorized into collective behavior, they may appear normal on an individual basis. 
As egocentric and collective behaviors emphasize different aspects, specific visualization designs should be introduced. It will be discussed when analyzing visualization techniques in the following sections.

\textbf{Data Types.}
A variety of data can be extracted from user behaviors across different domains. By analyzing multiple attributes of these data, we summarize four common data types including text, network, spatiotemporal information, and multidimensional data~\cite{jiang2016suspicious, chen2017social}. 
A brief explanation for each data types is described as follows. \textit{Text} provides semantic information of identities and backgrounds objects. \textit{Network}, also called subgraph, consists of a set of nodes interlinked with a set of edges. A formal definition of a graph can be found in~\cite{balakrishnan2012textbook}. \textit{Spatiotemporal information} captures spatial and temporal attributes of data. \textit{Multidimensional data} uses multiple attributes to describe the properties of objects. A detailed explanation of data types for each user behavior is introduced in the following sections.

\textbf{Anomaly Detection Techniques.}
The categorization of anomaly detection techniques used in this survey is borrowed from the survey written by Chandola~\etal~\cite{Chandola2009AnomalyDA}. The six categories are classification-based, nearest neighbor-based, clustering-based, statistical, information theoretic, and spectral anomaly detection techniques.
\textit{Classification-based} anomaly detection techniques develop models in the training phase and distinguish anomalies from normal data instances in the testing phase. In the training phase, classifiers are learned via training a set of data instances. In the testing phase, test instances are classified into one of the classes - normal or anomalous. 
\textit{Nearest neighbor-based} techniques compute anomaly scores from distance or relative density measures in a community. Anomalies are separated using distance-based nearest neighbor-based techniques, which calculate anomaly scores based on distance to its $k^{th}$ nearest neighbor. 
\textit{Clustering-based} techniques group similar data instances into clusters, and separate normal instances from anomalous instances.
\textit{Statistical} techniques presume probability distributions of data instances. Outliers are found in space of low probability whilst normal instances are observed with a high probability of occurrence. Statistical techniques can be further divided into parametric and non-parametric anomaly detection techniques providing whether there exists a model structure a priori.
\textit{Information theoretic} techniques analyze information content using measures such as entropy, relative entropy and Kolomogorov Complexity.
\textit{Spectral} techniques aim to find an approximation of the data by decomposing the problems and constructing suitable attributes. The attributes or components can then be embedded into lower dimensional subspace in which anomalous instances can be distinguished from normal instances.
A detailed explanation of categories and sub-categories can be referred to~\cite{Chandola2009AnomalyDA}. 

We focus our discussion on visualization works that apply anomaly detection techniques. A small proportion of visual analytic tools manage to detect anomalies by using carefully designed visualizations from which anomalous data instances can be visually distinguished from normal ones~\cite{cox1997brief, viegas2004digital,gatalsky2004interactive, weaver2007visual, erbacher2002intrusion}. The designs encode attributes and/or frequency using easily recognizable visual channels such as hues, heights of glyphs, sizes of nodes, etc~\cite{xiong1999peoplegarden, erbacher2002intrusion, viegas2004studying}. We exclude these papers in the discussion of anomaly detection techniques. 

\textbf{Visualization Techniques.}
We categorize visualization techniques that have been applied to anomalous user behaviors, including sequence, graph, text, geographic, chart, and glyph visualizations. \textit{Sequence visualization} illustrates relations between successive events with temporal information. Anomalous sequences include spreading patterns of rumors, sudden changes in the volume of posts, and unusual business processes. Common visual representations are timeline visualization, flow visualization, and parallel coordinates. 
\textit{Graph visualization} shows structured patterns composed of nodes and edges. Anomalous graph indicates special communication patterns in a group or communities, financial frauds conducted between employees and clients, and unauthorized network traces directed from sources to destinations. Typical graph visualizations are node-link diagram, circular-based designs (\ie a network topology map inside an outer ring), tree, and matrix. 
\textit{Text visualization} focuses on textual data. Anomalous text is indicated by specific keywords, topics, and sentiments extracted/abstracted from texts. Word cloud is one of the usual visualization techniques for text. Text can also be combined with other visualization techniques such as flow visualization to present more contextual information.
\textit{Geographic visualization} depicts mobility patterns of people or vehicles in geographic space. Mobility patterns include discrete as well as continuous patterns. Discrete patterns describe distribution and co-occurrence while continuous patterns depict trajectories of users when they move from one point to another. Abnormal mobility patterns are hot spots, an opposite traveling direction to most, and uncommon movement when compared to history. Heat maps and flows/bubbles projection on a geographic map are used most often for visual analysis of mobility patterns. 
\textit{Chart visualization} and \textit{Glyph visualization} represent the attributes of a multidimensional data item using a chart (\eg x-, y-axis, color of objects) and the feature of an icon (color, size, shape), respectively. Examples of anomalies include users who only reply in a discussion board but never initiate a post and who send an unusual amount of emails at a certain time.
Typical visualization techniques include 2D/3D scatter plot, bubble chart, bar chart, Gantt chart, etc. 

\textbf{Interaction Methods.}
Interaction plays an important role in visual analytics.
Based on analyzing interactions methods~\cite{yi2007toward} used in research works regarding detecting of anomalous user behaviors, we summarize the categories of interaction tasks including tracking \& monitoring, exploration \& navigation, knowledge externalization, pattern discovery, and refinement \& identification. 
Analysts may mark data of interest via click, hover or brush for \textit{tracking \& monitoring}.
Analysts may observe data via panning, zooming, or drill-down/roll-up functions for \textit{exploration \& navigation}.
Analysts may adjust attributes of data (\eg color, size, range) to reveal interesting patterns (\textit{pattern discovery}).
Analysts may collect, save, and extract the current visualization (\eg take a snapshot) for \textit{knowledge externalization}.
Analysts may label data with known identities (\ie abnormal or normal data item) for \textit{refinement \& identification} of results.

\section{Social Interaction}
\label{sec:social}
\textit{Social interaction} describes communication of ideas and thoughts between people. 
Social interaction can be further classified into private and public interaction. Private social interaction behaviors include sending and/or receiving emails, making phone calls, and sending text messages between familiars on a normal basis. Examples of anomalous interaction are communication of fraudsters~\cite{van2013reordering, van2014dynamic} and criminals~\cite{perer2006balancing, koven2018lessons}, emailing patterns of core contributors in a working group~\cite{fu2007visualization, gloor2004tecflow} and spam~\cite{muelder2007visualization}.
Public social interaction behaviors associate with posting/sharing/replying contents on publicly accessible social platforms. Specifically, writing reviews on e-commerce platforms and editing articles in Wikipedia are also counted as public social interaction. Anomalies related to this interaction consist of diffusion of rumors~\cite{zhao2014fluxflow, resnick2014rumorlens}, social bots~\cite{cao2016targetvue, shao2017spread}, and detection of events~\cite{marcus2011twitinfo, thom2012spatiotemporal, bosch2013scatterblogs2, pozdnoukhov2011space}. 

We observe a few differences between private and public social interactions. The linkage between senders and receivers is not explicit in public interaction compared to one-on-one conversations in private. The information accessible on public platforms is much more than that in private settings, leading to larger volumes of data collected relevant to public behaviors. The differences can also be implied from design principles of visual analytics tools which will be discussed in Section 4.3.

\subsection{Data Types}
Text data such as keywords, hashtags, and email contents help analysts comprehend social interaction behavior, as it provides information including sentiment, categories, and clusters of text under a certain topic. 
Gloor \etal~\cite{gloor2004tecflow}\cite{gloor2006identifying} filter emails by keywords that are known to be related to crime patterns. For example, ``bonus'' means the most important thing, ``investigation'' refers to what is coming up for criminals. 
TargetVue~\cite{cao2016targetvue} incorporates content features to detect social bot accounts. Mentioning of a topic under which sudden changes in the number of relevant tags are observed, is regarded as an anomalous behavior. 
Echeverria \etal~\cite{echeverria2017discovery} discover a bot network in Twitter by solely mining the textual features of tweets. They found that the tweets of the botnet are taken directly from ``Star Wars'' novels. 
Beagle~\cite{koven2018lessons} allows analysts to filter contents from a filter set as well as to construct filters using keywords that are found useful during the investigation of scamming activities. 

As social interaction concerns with passing, sharing, and exchanging information, network are often seen when conversations are held between users. Follower relationship in social media, back-and-forth communication via emails, and amendments made by one user in Wikipedia in response to the edit of another user are considered as network data. 
Gloor \etal~\cite{gloor2004tecflow} identify the team leader, practice leader, and practice coordinator from visualization of social email networks. These anomalous users are placed in the center of the social network and connected to multiple nodes. 
Fu \etal~\cite{fu2007visualization} explore small-scale email networks, where a node represents an email address, and an edge between two nodes indicates an email exchange. Analysts are able to identify different email networks for specific research groups as little communication is made across different groups. FluxFlow~\cite{zhao2014fluxflow} derives user networks when exploring the process of anomalous information spreading. Indegree and outdegree are extracted based on the interaction graph of a Twitter user. These measures signal the influential power of the user. 

Temporal information can be found from timestamps of microblogs, time and date of emails and calls, and days when a user appear on a forum. Location of geo-located microblogs, the location of calls, and the terrorist network of a country are spatial data. Temporal data facilitates the analysis of communication evolution whereas and spatial data explains where the behavior occurs. 
Elzen \etal~\cite{van2013reordering, van2014dynamic} detect communication bursts using dynamic network visualization. One important part is the temporal analysis of events (\eg mobile phone calls), where trends opposite to global trends, periodic repetition, and a sudden block between homogeneous behaviors are considered abnormal. 
CloudLines~\cite{krstajic2011cloudlines} regards sudden changes in the number of specific keywords within a period as anomalies. The keywords are collected from tweets, which arrive in data streams at non-uniform time intervals. 
Some visualization works combine temporal and spatial analysis in event detection. ScatterBlogs~\cite{thom2012spatiotemporal, chae2012spatiotemporal} detects events containing geographic information such as power outages and disasters from microblogs, and in the meantime represent messages related to the events on a map. 

Multidimensional data for detecting anomalous user behaviors include the length of a tweet, number of posts/emails, and average rating scores in e-commerce platforms. Multidimensional data not only offers comprehensive descriptions of social interaction, but also helps abstract anomalousness of behaviors. 
Webga and Lu~\cite{webga2015discovery} detect anomalous ratings by incorporating multidimensional data into the analysis. The multidimensional data includes the scores given by every user at the corresponding time. Rating frauds are discovered by measuring differences in average ratings and the number of rating activities in two time windows. 
Cao \etal~\cite{cao2016targetvue} detect anomalous users in social media by carefully selecting communication features. To investigate the interaction aspect of a social account, features such as whether users tend to communicate within a group or spread information in public, and whether users are responded from others are measured. 
FraudVis~\cite{sun2018fraudvis} selects ten features based on the rank of anomaly score to investigate which features contribute most to frauds on the Internet. The activity count within different time periods, for instance, is one of the features that evaluate the number of views on a video website. 

\begin{figure*}[!t]
    \centering
    \includegraphics[width=\textwidth]{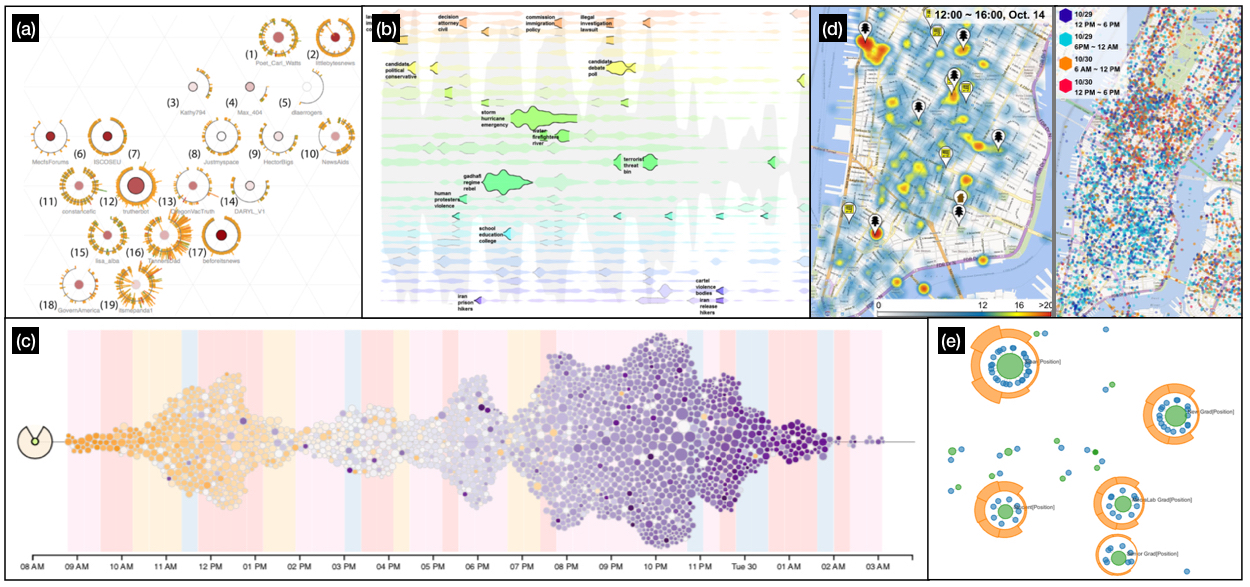}
    \caption{Visualizations of anomalous social interaction behaviors. (a) TargetVue~\cite{cao2016targetvue} uses circle-based glyph visualization to encode individual users' temporal posting/reposting behaviors, anomalousness of their behaviors, and correlation between suspicious users. (b) Leadline~\cite{dou2012leadline} visualizes event episodes using horizontal pulse-shaped timeline visualization. (c) FluxFlow~\cite{zhao2014fluxflow} shows anomalous information spreading on social media using packed circle timeline visualization. (d) Chae \etal~\cite{chae2014public} present public behavior responses to disaster events in microblog using a heat map and hexagons on a map. (e) Mobivis~\cite{shen2008mobivis} visualizes the calling behavior of a network consisting of university staff and students using a node-link diagram.}
    \label{fig:social} 
\end{figure*}

\subsection{Anomaly Detection Techniques}
Classification-based techniques are popular in discovering abnormal social interaction when compared to the application of the techniques in the other three user behaviors. 
The retrieval of ``Star Wars'' botnet~\cite{echeverria2017discovery} is achieved with a na\"ive Bayesian classifier based solely on textual features. This basic technique is effective because the tweets posted by the botnet are cited from the ``Star Wars'' novels. 
ScatterBlogs2~\cite{bosch2013scatterblogs2} proposed a supervised, Support Vector Machine (SVM) classification-based approach to train classifiers as user-adjustable filters. 
A random forest algorithm, \ie rule-based classification detects misinformation that is spread by social bots in a supervised approach \cite{shao2017spread}. 
RumourLens~\cite{resnick2014rumorlens} analyzes the impact of rumors during the information diffusion process. It performs iterative expansion of a query set and iterative refinements of a classifier (ReQ-ReC retriever and classifier)~\cite{li2014req}. The output is a ranked list of tweet clusters that seem to be rumors, which can be refined by users. 
FluxFlow \cite{zhao2014fluxflow} utilizes one-class conditional random fields (OCCRF) to perform sequential anomaly detection. The OCCRF model assumes the highly dynamic and one-class nature of anomalies, and computes an anomaly score by measuring dissimilarity from unlabeled training samples. The dissimilarity is derived from the difference of the posterior probability of a normal label and that of an abnormal label. 

Nearest neighbor-based techniques calculate anomaly scores from distance or density. 
Metrics such as density, betweenness centrality, and group degree centrality in networks are the ranking criteria of homogeneity/risk for Collaborative Innovation Networks~\cite{gloor2003visualization}. 
MobiVis~\cite{shen2008mobivis} incorporates semantic information of phone calls and geographic proximity into a heterogeneous graph. Through importance filtering based on variables such as node degree of a neighborhood, important nodes and edges can be pruned from interaction with the ontology graph. 
TargetVue~\cite{cao2016targetvue} employs time-adaptive local outlier factor model to quantify sudden changes of posting or emailing behaviors. A user can be identified as a time-series vector in multidimensional feature space. Each user is given an anomaly score computed from features that distinguish one user from others, and from his/her own history. 
Kernel density estimation (KDE) is used for computing continuous distribution. It scales the parameters of estimation by enabling the kernel scale to vary based on the distance from the point to the $k^{th}$ nearest neighbor in a data set. Cloudlines~\cite{krstajic2011cloudlines} allows logarithmic distortion of amplifying recent events in time. A kernel density estimator and a truncation function help focus on recent events that appear dense in time series. KDE is also used in~\cite{pozdnoukhov2011space} to inspect spatiotemporal regularities of topics. Point patterns are related to continuous regions by comparative kernel density analysis.

Statistical techniques are used in event detection, where anomalousness is quantified by measuring differences from models constructed from history behaviors. 
TwitInfo~\cite{marcus2011twitinfo} finds peaks from time-series events by considering exponentially weighted moving average and variance in a time window. The algorithm starts a new window if a significant increase in counts relative to the historical mean is encountered.  
Iterartive non-parametric regression based on Loess smoothing decomposes time series of interest to three components: trend, seasonal, and remainder component. Z-scores of remainder values are abnormality rating. This novel method was first used in ScatterBlogs \cite{chae2012spatiotemporal}. It was later applied to identify unusual topics in the selected regions \cite{chae2014public} and used as part of predictive analytics based on topic trends in historic time series \cite{yeon2017predictive}.

EventRiver~\cite{luo2012eventriver} applies a clustering-based approach based on temporal locality in the analysis of streaming texts. The clusters are related in contents regardless of time spans. 
ScatterBlogs~\cite{thom2012spatiotemporal} uses the Lyold clustering technique to distinguish unusual events from general message clusters originating from high densities in time and space. 
Episogram~\cite{cao2016episogram} select appropriate features for clustering, and generate clusters that are always centered at the positions with highest densities in the data space. 
FraudVis~\cite{sun2018fraudvis} employs the CopyCatch algorithm, a graph-based clustering approach to explore fraud groups who suddenly follow a user in social media on a single day. 

Spectral techniques are used to detect interesting network structure of editing histories~\cite{brandes2009network} and rating frauds in e-commerce systems~\cite{webga2015discovery}. 
Brandes \etal~\cite{brandes2009network} abstract weighted attributes on nodes and edges from users and relationships between users respectively. A weighted graph is projected into controversy space where collaboration or competition structure of two user groups are easily identified. 
Webga \etal \cite{webga2015discovery} adopt a dimension reduction algorithm, singular value decomposition (SVD), to detect fake ratings that are written to boost the popularity of selected items in e-commerce stores. Once the suspicion level is raised above a threshold, alerts are sent to the visualization.

\subsection{Visualization Techniques}
\textit{Egocentric Behaviors.} Egocentric Social Interaction behaviors study the role of a user from his/her interaction with others. Examples of anomalous egocentric behaviors are users who only reply in a discussion board or who send an unusual amount of emails at a certain time. 
We observe that glyph, text, and graph visualizations are favored visual representations for egocentric behaviors. 

Anomalous user behaviors can be identified via glyph visualization that are in different appearances to those of normal ones.  
Episogram~\cite{cao2016episogram} uses arrow-based and arc-based timelines to demonstrate posting and reposting activities, respectively. The two timelines can be aggregated to obtain overall tweeting behaviors. Users who always repost immediately after a message is posted are identified as arcs that always start from one end. 
TargetVue~\cite{cao2016targetvue} (Figure~\ref{fig:social}~(a)) tackles the challenge of discovering social bots in Twitter. The circle-based glyph visualization facilitates investigation in terms of topics, sentiments, temporal dynamics of communication and its impacts, and relationship among accounts. Specifically, individual users' temporal posting/reposting behaviors, anomalousness of their behaviors, and the correlation between suspicious users are encoded by behavior glyph, feature glyph, and relation glyph, respectively. 

Text visualization can be used to describe egocentric communication patterns in emails~\cite{viegas2004digital, viegas2006visualizing}. 
PostHistory~\cite{viegas2004digital} shows the evolution of emailing patterns. It consists of two views, with one revealing the intensity of exchanged messages with each contact in a calendar view, and the other demonstrating how email addresses evolve over time in movies. Analysts can change addresses' positions by vertical/circular/alphabetical arrangement.
Social Network Fragment~\cite{viegas2004digital} represents social networks in a graph where nodes are replaced by colored names of individuals. The larger the font of the name, the stronger an individual is tied to others. 
Vi\'egas \etal~\cite{viegas2006visualizing} study changes of relationships implied from changes of keywords in email contents. The frequency and distinctiveness of keywords can be inferred from the sizes of texts, and thus anomalies such as changes of relationships (\eg from peer to boss) can be inferred. 

In addition to glyph and text visualization, graph visualization, especially node-link visualization helps detect anomalous individual behaviors from their social interaction.  
Li \etal~\cite{li2004email} explore email patterns in two graphical modes: cliques and email flows. A spam bot is detected in the email flow panel when only edges originating from one node are visualized.
Gloor \etal~\cite{gloor2003visualization, gloor2004tecflow, gloor2006identifying} investigate communication patterns of working groups in node-link visualization, and study the evolution of social structures over time in animation. Networks are drawn in personalized mode or subject mode to identify core contributors in groups and important messages, respectively~\cite{gloor2003visualization}.
The visualization tool TeCFlow~\cite{gloor2004tecflow, gloor2006identifying} detects the hidden communication structure from the Enron email corpus. The hierarchical social networks uncover how Enron employees conduct collusion and frauds by emphasizing the roles of influencers, gatekeepers, and leaders. Semantic node-link views enable investigation in terms of email addresses, keywords or time. 
Shao \etal~\cite{shao2017spread} evaluate the extent to which an account expresses similarity to the characteristics of social bots based on diffusion patterns of tweets. In the ``Hoaxy'' platform, a node-link diagram represents the social networks, with brighter hues indicating higher anomalous scores.

\textit{Collective Behaviors.} Collective Social Interaction behaviors derived from users acting in a group or acting in response to each other. 
Anomalous collective social behaviors include temporal development of tweets, the reaction of people to special incidents, and separate group patterns of communication. Sequence, geographic, and graph visualizations used often for collective behaviors. 

Sequence visualization represents the evolution of collective behaviors in various forms such as parallel coordinates and pulses/bubbles arranged along a timeline visualization. 
Vi\'egas \etal~\cite{viegas2004studying} visualize revision history of Wikipedia pages in modified parallel coordinates. Each revised version of an article is represented by a vertical axis, with the axis' length indicating the length of the article. The vertical axis is divided into parts with each corresponding to revisions made by every author. By linking the axes together, a modified form of parallel coordinates shows the competition/mass deletion histories of articles. 
RumorLens~\cite{resnick2014rumorlens} demonstrates the movement between different states of interaction with a rumor. The main view shows a Sankey diagram. The number of people exposed to the rumor and the associated correction is illustrated with lengths of colored segments (blue for rumors and red for corrections) in one axis. By linking different states between axes that correspond to time epochs, analysts can understand the influence of rumors and the corrections. 

Pulses and bubbles arranged according to temporal sequence illustrate the anomalies of collective social interaction behaviors. Major changes in temporal development of texts are detected by highlighting unusual shapes of timelines. 
As one of the earliest visualizations that investigate the emergence of events, TwitInfo~\cite{marcus2011twitinfo} visualizes bursts of events in a line chart. The highlighted and labeled event peaks suggest events that trigger heated discussion on Twitter. 
CloudLines, LeadLine and EventRiver~\cite{krstajic2011cloudlines, dou2012leadline, luo2012eventriver} detect events by relating volume of text data extracted from online news within a period of time to temporal density of keywords. Horizontal pulse-shaped timeline visualization represents event episodes, with the sizes of pulses indicating the importance of events. 
LeadLine (Figure~\ref{fig:social}~(b)) and EventRiver~\cite{dou2012leadline, luo2012eventriver} arrange vertical positions of events according to similarity of topics. 
FluxFlow~\cite{zhao2014fluxflow} (Figure~\ref{fig:social}~(c)) discovers temporal trends and impacts of users in information spreading process (\eg rumors). The main view consists of packed circles arranged along a timeline. A user's influence (\ie the number of followers) and anomaly score are encoded by the size and color of a circle, respectively. A user can be analyzed from three perspectives simultaneously: tweet volume, sequence, and distribution of anomalous accounts. A complementary tree visualization demonstrates the correlation of user accounts in the diffusion process. 

Geographic visualization is used to reveal events containing spatial as well as temporal references. With geographic details, anomalies can be detected from spatial intensities obtained from a collection of social interaction behaviors.  
Lee \etal~\cite{lee2010measuring} introduce one of the earliest works of applying spatiotemporal analysis to social media, where flows of people are represented as arrows on a map. 
ScatterBlogs~\cite{thom2012spatiotemporal, chae2012spatiotemporal} employ geographic visualization for anomaly detection of topics and events as well as their spatial and temporal marks. 
ScatterBlogs2~\cite{bosch2013scatterblogs2} uses dots on a map to portray geo-located microblog posts. It differs from its previous version since there are two settings in ScatterBlogs2: a classifier creation environment and a monitoring environment. Analysts create task-tailored filters based on messages of well-understood events in the classifier creation environment, and obtain contexts of interesting events from a filter orchestration view and a time slider in the monitoring environment. 
Thom \etal~\cite{thom2012spatiotemporal} extract terms from messages and cluster topics as tag clouds on a zoomable map. Anomalous events are labeled and positioned on a map according to its detected location.
The ``Star Wars'' botnet was discovered by accident when Echeverria \etal~\cite{echeverria2017discovery} observed sharp boundaries of the latitudinal and longitudinal position of some tweets, which were generated from bots considering the unusual spatial distribution. 

Heat map, one of the geographic visualizations, is effective at illustrating geographically-marked microblog messages. 
Pozdnoukhov \etal~\cite{pozdnoukhov2011space} compute heat maps from streaming tweets. Density of heat maps indicates spatial variability of population's response to various stimuli such as large scale sportive, political or cultural events. The difference in density between two heat maps implies temporal evolution of events. 
Chae \etal~\cite{chae2014public} (Figure~\ref{sec:social}~(d)) collect a sheer volume of real-time microblog messages and mine public behavior response to disasters. A heat map and hexagons on a map identify spatiotemporal differences between crisis and normal situations. 

Graph visualization including node-link and circular-based visualization uncover anomalous structures of social interaction. 
Perer and Shneiderman~\cite{perer2006balancing} emphasize the need to examine social networks systematically in \textit{SocialAction}. The visualization tool is designed accordingly to encourage interaction with clustered node-link visualization. Analysts can quickly direct their attention to the most anomalous networks as nodes/subgroups are colored according to their ranks of anomalousness. 
Fu \etal~\cite{fu2007visualization} examine small-world email networks using several visualizations. For example, stacked displays of graphs on a spherical surface visualize communication patterns between different groups. A hierarchical drawing emphasizes important nodes by placing them high in the hierarchy. 
MobiVis~\cite{shen2008mobivis} (Figure~\ref{fig:social}~(e)) visualizes the calling behavior of a network consisting of university staff and students using a node-link diagram. The goal is to investigate information exchanges and the implicit social relationship. The researchers design a ``behavior ring'' for user(s), which arrange events in a radial form around a node. Analysts study structural interaction from the correlation between nodes and temporal interaction from the rings. 

Circular-based representation demonstrates collective social interaction behaviors in a packed visualization. 
Elzen \etal~\cite{van2013reordering, van2014dynamic} combine the circular hierarchical edge bundle view and massive sequence view (MSV) to detect unexpected suspicious communication patterns. The novelty of this visualization tool is that it incorporates node reordering strategies in MSV. The reordering techniques take account of closure, proximity, and similarity to ensure outliers stand out from mass data. 
Webga and Lu~\cite{webga2015discovery} project nodes (\ie users) into a circular layout to discover rating frauds from the temporal relationship between users and items. The combination of singular value decomposition diagram, re-ordered matrix representation, and the temporal view reveals interesting group patterns of items. These patterns share a similar rating history and users of similar behaviors.

\subsection{Interaction Methods}
Visual analytics of social interaction behaviors applies tracking \& monitoring as one of the first steps of exploratory analysis.
TwitInfo~\cite{marcus2011twitinfo} tracks bursts of events in time series by highlighting the event peaks in a line chart. These peaks suggest events that trigger heated discussion on Twitter. 
Koven \etal~\cite{koven2018lessons} multi-select summaries of email contents in the main panel to keep track of important keywords regarding scamming activities. 
FluxFlow \cite{zhao2014fluxflow} monitors information diffusion using multiple coordinated views. As analysts select a point in tree view, the diffusion pattern generated by the user's reposting behavior is shown in the thread view. The interaction is usually achieved in tools with multiple coordinated views~\cite{van2013reordering, viegas2004digital, shen2008mobivis, cao2016targetvue, morstatter2013understanding, viegas2004newsgroup}.

Exploration \& navigation allows analysts to focus on different subranges of data flexibly. 
V\'egas \etal~\cite{viegas2006visualizing} design a scrolling bar, allowing analysts to review email conversation in different periods of time. 
TargetVue~\cite{cao2016targetvue} enables analysts to zoom and pan in global and inspection view to locate to anomalous areas.
Exploration in Episogram~\cite{cao2016episogram} is not limited to zooming function. Analysts can select a user of interest, and aggregate all users who perform the same posting/reposting activity. In this way, an individual's details as well as the general trend are obtained. 
MobiVis~\cite{shen2008mobivis} designs a ``behavior ring'', from which analysts select different levels of granularity to arrange calling events in a radial form around a node. The length of petals corresponds to the duration of selected events.

Pattern discovery is achieved in various forms of interaction such as filtering. 
Gloor \etal~\cite{gloor2003visualization} visualize email data to discern the structure of networks and identify core contributors. Emails are presented according to the type of links (\ie ``To''/``From''/``Cc'') in the email network. 
ScatterBlogs2~\cite{bosch2013scatterblogs2} supports generation of task-tailored filters in the classifier creation environment. In the monitoring setting, analysts can orchestrate the filters to detect anomalous users. 
Sorting visual objects also uncovers interesting patterns.
Cloudlines~\cite{krstajic2011cloudlines} visualizes online news events in timelines in either linear or logarithmic scale. The tool allows analysts to reconfigure visual objects via click and drag. 
Webga and Lu~\cite{webga2015discovery} detect rating frauds in the projection view, which contains two orthogonal axes inside a circle. Analysts can choose any two dimensions and the mapping method to dig out the outlier pattern.
Changing encoding scheme is useful. 
Chae \etal~\cite{chae2014public} demonstrate events detected from microblog messages with a heat map, scatters, and hexagons on a map. 
TargetVue~\cite{cao2016targetvue} encodes users' action in a time sequence, anomalousness of their behaviors, and correlation to three glyph designs, so that analysts acquire various perspectives of the social accounts.

Analysts may want to save results of analysis for future study. 
For example, documents of interest can be saved in the evidence box of the EventRiver~\cite{luo2012eventriver} visualization tool. This function supports hypothesis evaluation and evidence exchange. 
Koven \etal~\cite{koven2018lessons} allow analysts to share tags created during analysis of email contents. 
Visualization on a website tends to have more flexible applications of knowledge externalization than stand-alone tools. 
After one analyze the anomalous extent of social bots in the Hoaxy platform (https://hoaxy.iuni.iu.edu/)~\cite{shao2017spread}, the results can be saved into CSV files for sharing.

Refinement \& identification is conducted after analysts have obtained a basic understanding of social interaction behaviors. 
LeadLine~\cite{dou2012leadline} associates events with corresponding time-sensitive keywords automatically. Analysts can then annotate the events manually to provide accurate labels. 
There are two labeling strategies in EventRiver~\cite{luo2012eventriver}: representative event labeling and outlier labeling. On one hand, representative labeling is for events that contribute to the biggest cluster of a story. On the other hand, outlier labeling labels outlier events in a story. 
Koven \etal~\cite{koven2018lessons} emphasize tagging abilities in discoveries of anomalies. Analysts can label an account as a scammer, victim, service, or other categories. These tags can be used for creating filters as well as the calculation of statistics about scamming activities.

\section{Travel}
\label{sec:travel}
\textit{Travel} is physical movements of users between places containing geographic information. Analysis of travel behaviors is meaningful for traffic monitoring, urban safety, and urban planning~\cite{cao2018voila}. 
Travel behavior data can be collected from mobile phones and base stations, Global Positioning System (GPS), maritime search and rescue events, and medical records. 
Anomalous travel behaviors differ from the expected patterns indicated by individual historic records or activities of the crowd. Examples include irregular driving direction~\cite{liao2010anomaly, cao2018voila}, hotspots (\eg crowded neighborhoods) \cite{cao2018voila, wu2016telcovis, ferreira2013visual}, and characteristic travel patterns associated with groups of travelers~\cite{weaver2007visual, beecham2014characterising}.
These anomalous behaviors can reveal potentially harmful events such as disease outbreaks and terrorist attacks.

\subsection{Data Types}
Spatiotemporal data is essential to describe the information of when and where about users' physical motion. 
Spatial data consists of latitudes and longitudes, trajectories, pickup/drop-off locations, locations of base stations, etc. Temporal data includes timestamps of indoor activities, estimated time arrival, and pickup/drop-off date and time. Analysis of travel behavior usually combines both spatial and temporal data. 
Pu \etal~\cite{pu2011visual} explore mobility patterns of different user groups from mobile phone data collected from each base station and handoff data (\ie successive calls with different base station IDs). Spatiotemporal data related to communication include the start time of calls, time duration, the city of the opposite side of calls, and location and direction of base stations. 
TelCoVis~\cite{wu2016telcovis} explores co-occurrence of people using telco data, which is a type of all-in-one mobile phone data containing activity records of calls, messages, and Internet usage. Data of each type of activity is comprised of timestamps, base station ID and the corresponding latitude and longitude. 
Kim \etal~\cite{kim2018data} create a visualization that helps comprehend flow patterns by analyzing the spatial distribution of non-directional discrete events over time. 

Multidimensional data enriches skeletons of analysis of travel behavior. A combination of attributes including distance traveled, speed of cars, tip amount and toll amount for taxi trips, and frequency of residents' indoor activities provides a detailed description of travelers or vehicles. 
Pu \etal~\cite{pu2011visual} aggregate multidimensional data associated with base stations and mobile phone users. The data includes the total number of phone calls made by each user at each station and at all stations, in addition to spatiotemporal details.
Malik \etal~\cite{malik2011visual} evaluate the potential risks of Coast Guard search and rescue (SAR) operations to better plan response actions to mitigate risks. The SAR data consists of two components: response cases and response sorties. Multidimensional data of each component contains the number of lives saved, lost, and assisted. 
Voila~\cite{cao2018voila} extracts multidimensional features to detect abnormal incoming and outgoing taxi flows in a cell (a region is segmented into multiple cells). Examples of the features are the number of vehicles that flow in and out from one cell to another. Analysis of inflows and outflows for multiple cells consist of multidimensional data. 

Text associated with travel behavior is mainly used for identification and categorization. Examples include user ID, textual messages, and roam type and toll type. 
Pu \etal~\cite{pu2011visual} collect information of mobile phone ID, International Mobile Equipment Identity, city ID, roam city, roam type, and toll type to describe properties of mobile phones. These details help explain the nature of mobile phone users, \ie travelers. 
Beecham \etal~\cite{beecham2014characterising} categorize people into different groups in order to summarize group-cycling behaviors. Cyclists under the cycle hire scheme are classified according to age, sex, full postcode, whether they cycle more with others or on an individual basis, and spatiotemporal information. 
Liao \etal~\cite{liao2017visual} study resident indoor activities. These activities include not only long-term activities such as sleep, relax, watch TV, but also short-term ones such as entering home. 

Network data refers to trajectories between origins and destinations. Network data is mainly used to complement spatiotemporal analysis. 
Ko \etal~\cite{ko2014analyzing} assess flight journeys that often delay by analyzing pairs of origin and destination airports. By aggregating the amount of delays for each flight journey (\ie network), analysts detect anomalous airports and flights where prevalent delays are often found. 
Beecham \etal \cite{beecham2014characterising} study group-cycle journeys that link starting points and destinations.

\begin{figure*}[!t]
    \centering
    \includegraphics[width=\textwidth]{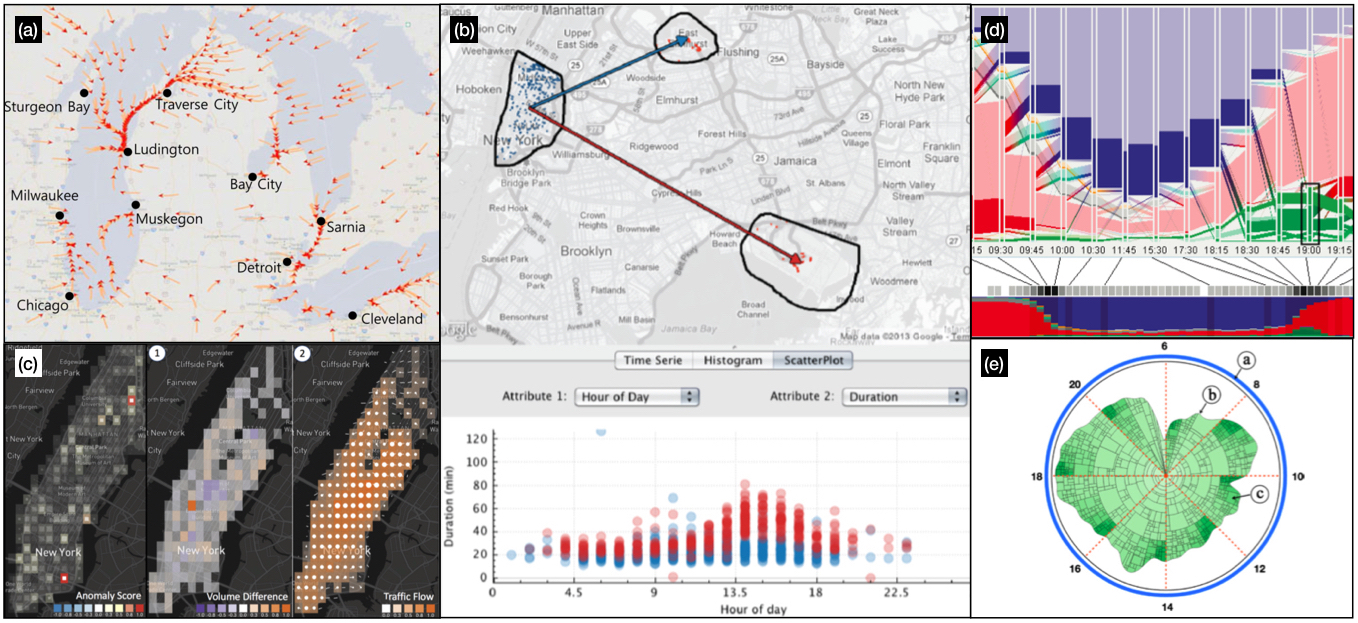}
    \caption{Visualizations of anomalous travel behaviors. (a) Kim \etal~\cite{kim2018data} show origin and destination via directions of arrows in a flow map. (b) Ferreira \etal~\cite{ferreira2013visual} investigate anomalous taxi trips in New York city in multiple coordinated views of a dot map and a line chart. (c) Voila~\cite{cao2018voila} displays unusual traffic flows between a focal region using heap map. (d) Von \etal~\cite{von2012visual} visualize different types spatiotemporal patterns by parallel coordinates. (e) Wu \etal~\cite{wu2016telcovis} design a contour-based treemap to illustrate spatial and temporal characteristics of human mobility at a specific place.}
    \label{fig:travel} 
\end{figure*}

\subsection{Anomaly Detection Techniques}
Statistical anomaly detection technique is the most often used to analyze travel behaviors.  
A data-driven approach~\cite{riveiro2009interactive} using self-organizing maps and Gaussian mixture models are applied to describing normal behaviors of vessels. By comparing data with rules and signatures, unusual traveling patterns are detected through visual analytics. 
A box-plot method~\cite{lee2010measuring} checks whether geographical regularity deviates from normal conditions by large extents. 
Two visualization works~\cite{maciejewski2010visual, von2012visual} apply cumulative sum (CUSUM) algorithms after kernel density estimation to better identify outliers in time series. One \cite{maciejewski2010visual} calculates density estimation for the event category as well as density estimation for all categories, and obtain the expected number of events within a given area. Outbreaks in the temporal domain can be detected with the cumulative summation algorithm for the given location. Applying CUSUM after kernel density estimation enables analysts to spot spatial areas worth investigation quickly, and then analyze historical time series to look for unusual trends. The other work \cite{von2012visual} also utilizes CUSUM algorithm to trace uncommon development patterns.

Clustering-based is employed to reduce computation complexity and visual clutter for large-scale databases. 
Andrienko \etal~\cite{andrienko2013visual} use k-means clustering to analyze spatiotemporal phenomena described by multiple spatial time series. The clustering approach groups spatial objects by the similarity of their corresponding time series, and thus spatially unusual events can be detected. The clustering approach is used in conjunction with statistical methods to model time series such that residuals are randomly distributed over time. High deviations from expected time values are seen as anomalies. K-means clustering is also used to detect anomalies of mobility patterns around base stations~\cite{pu2011visual} and group-cycling behavior~\cite{beecham2014characterising}. This clustering approach requires the number of output clusters to be specified before computation.
Lin \etal~\cite{lin2005visualizing} propose VizTree and Diff-Tree to mine anomalous patterns by comparing time series (\eg yoga postures) with normal references. It uses bottom-up hierarchical clustering to produce a nested hierarchy of similar groups of objects based on a pairwise distance matrix. 
TelCoVis~\cite{wu2016telcovis} applies a biclustering technique in binary matrices, where 1 means co-occurs of human mobility and 0 means otherwise. Thus, origins and destinations of human mobility can be bundled into coordinated sets as biclusters. 

Nearest neighbor-based anomaly detection techniques compute the continuous distribution for detection and anomaly scores. 
KDE computes the spatial and/or temporal distribution of discrete events, which is particularly useful for detecting hotspots in density-based visualizations. Malik \etal \cite{malik2011visual} employs a modified variable KDE technique to identify spatial hotspots of search and rescue cases in the U.S. Coast Guard. 
Kim \etal \cite{kim2018data} compute continuous spatiotemporal distributions of discrete events by applying the KDE approach to two-dimensional data, which is achieved without trajectory information. 
Local outlier factor (LOF), a density-based nearest neighbor-based technique is used to calculate anomaly degree of indoor daily activities of residents. Duration, number of times, and start time are selected as the properties to compute outliers. 


\subsection{Visualization Techniques}
\textit{Egocentric Behaviors.} Egocentric Travel Behavior is individual physical movement in geographic space. An example of anomalies associated with egocentric travel behavior is an unexpected increase in time spent on indoor activities. Chart visualization is seen to represent egocentric travel behaviors. 

VizTree~\cite{lin2005visualizing} uses suffix tree visualization to indicate abnormal parts of the time series by comparing with reference (\ie normal) patterns. Anomaly detection is achieved by transforming a time series into a symbolic representation and visualizing it as a modified suffix tree. 
Weaver \etal~\cite{weaver2007visual} explore individual hotel visitors in a calendar view, a map view, and an arc diagram. A calendar view shows total visits on each day, with squares and circles indicating weekends and weekdays, respectively. A multi-layer map view describes paths from residences to hotels, relative to railroads and rivers. By synthesizing temporal and spatial patterns observed from multiple views, analysts obtain circuitous routes taken by salesmen, cooperation between traveling merchants, and the effects of weather and seasonal variations, etc. 
Liao \etal~\cite{liao2017visual} are interested in resident behaviors recorded by smart home visual systems. A heat Gantt chart view shows start time, duration, and the number of occurrence of different activities on a daily basis. By combining the heat Gantt chart with other views, activities that deviate from daily routines are detected through comparison on different daily records.

Geographic visualization is also seen for egocentric travel behaviors. A transit map displays GPS traces~\cite{liao2010anomaly} of moving taxis in basic mode, monitoring mode, and tagging mode. Taxis are represented by glyphs on the map, with the colors dependent on whether the taxi is loaded with passengers. A taxi with an irregular driving direction or moving at high speed, and a crowded neighborhood are egocentric anomalous travel behaviors. 

\textit{Collective Behaviors.} A collection of users move together in time and space, we say their travel behaviors are collective. Abnormal travel behaviors can be identified from regions crowded with people. 
As most visualization tools studying collective travel behaviors employ geographic visualization, we analyze travel behaviors using the finer categories under geographic visualization including flow maps, heat maps, and bubble/dot map. 

Flow maps represent trajectories by linking origins and destinations on a map. 
Andrienko~\cite{andrienko2013visual} proposes a framework for spatiotemporal analysis and modeling. Anomalies are found in temporal line charts displaying model residuals. Spatial flows between cells are represented by directed half-arrows whose widths are proportional to the total counts of objects that move. The flows are laid upon Voronoi maps. 
Trajectories of cycling patterns are shown as flows on a London city map~\cite{beecham2014characterising}. The straight and curved end of a flow represent origin and destination, respectively. Group journeys are colored red on the map whereas non-group journeys are colored blue. One of the findings is that female cyclists are more likely to make late evening journeys when cycling in groups.
Kim \etal~\cite{kim2018data} (Figure~\ref{fig:travel}~(a)) extract, represent, and analyze flow maps and heat maps of spatiotemporal data without the use of trajectory information. The flow map visualizes origin and destination via directions of arrows, and the difference of flows are encoded in heat maps. Hot spots can be found with this visualization. 

Heat maps display spatial densities of collective travel behaviors. 
Maciejewski \etal~\cite{maciejewski2010visual} develop an interactive visual environment to dig out hot spots in spatiotemporal data for crime analysis or surveillance syndrome. Bivariate and multivariate heat maps help detect spatiotemporal hot spots by combining height maps, colors, and contours. 
To analyze risks of Coast Guard search and rescue (SAR), Malik \etal~\cite{malik2011visual} identify potential hot spots using heat maps. Risks of stations are indicated by the intensity of colors. The red heat map shows the time taken by stations to deploy an asset to an SAR accident while the green heat map indicates the SAR coverage.
Ferreira \etal~\cite{ferreira2013visual} (Figure~\ref{fig:travel}~(b)) investigate anomalous taxi trips in New York city in multiple coordinated views of a dot map and chart visualizations. Dots on a map imply pickup and dropoff sites in the region. In the cases of Hurricane Sandy and Irene, there are virtually no dots during hurricanes, but traffic seemed to go back to normal in the following days. 
Voila~\cite{cao2018voila} (Figure~\ref{fig:travel}~(c)) explores taxi trips to detect sudden changes in traffic patterns. There is an anomaly detection mode giving visual cues of regional anomalies, and a context mode providing information of volume difference, traffic flow, and expected patterns at different times. Unusual traffic flows between a focal region and two other places are highlighted by the deep red color of heat maps. Feedback from analysts can update the anomaly score and thus change the color of heat maps for the selected region. 

We analyze other visualization techniques for travel behaviors including sequence and graph visualization. 
Von \etal~\cite{von2012visual} (Figure~\ref{fig:travel}~(d)) categorize spatiotemporal patterns into different types of locations according to home, work, tennis, etc. The main view is Dynamic Categorical Data View in a varied form of parallel coordinates, which show the evolution of all types of data. Each axis of parallel coordinates indicates a point in time. When analysts select a type of data, related geographic information is plotted in the linked map, where arrows on the map indicate physical movement of people. 
In TelCoVis~\cite{wu2016telcovis}, Wu \etal design a contour-based treemap to illustrate the spatial and temporal characteristics of human mobility. By combining with heat map, matrix, and parallel coordinates, analysts gain insights into co-occurrence of human mobility and correlations of co-occurrence.

\subsection{Interaction Methods}
Analysts track and monitor data to look for anomalies. 
Uninteresting and expected patterns can be unmarked~\cite{lin2005visualizing}. This improves the efficiency of detection processes and reduces false positives. 
TelCoVis~\cite{wu2016telcovis} emphasizes the correlation between spatial and temporal data for exploring the co-occurrence of human mobility. When analysts hover on a sector in the contour-based treemap, all sectors corresponding to the same region will be highlighted. Moreover, analysts can mark the region for exploration. 
Analysts can track a set of features of categoric data~\cite{von2012visual} including location, movement pattern, group membership, and group changes. The selected data instances are highlighted in the linked map view and the categoric view.

The interactions associated with exploration \& navigation piece separate fragments of data. 
Panning and altering views via scrollbars facilitate detection of non-trivial patterns in large time series databases~\cite{lin2005visualizing}.
High-level outlooks and details should be accessed interchangeably when exploring travel behaviors. Different levels of aggregation in time~\cite{malik2011visual, ferreira2013visual, andrienko2013visual} and space~\cite{ferreira2013visual, ko2014analyzing, cao2018voila, beecham2014characterising} are seen in a variety of visualization tools.

Unusual travel patterns are uncovered by filtering, configuration, and encoding to various visual forms. 
The anomaly grading view in SHVis~\cite{liao2017visual} present anomaly scores of selected activities. Analysts click on different days and drag date intervals to compare the activities during the different periods of time. 
In order to analyze maritime operations and assess risks associated with the allocation of resources~\cite{malik2011visual}, analysts generate a combination of filters which can be applied to spatial regions and temporal plots. In addition, analysts can evaluate the effects as a result of opening/closing a station, and determine which station is suited for closing. 
Visualization can be altered in color and in form to reveal anomalous patterns. 
Andrienko~\cite{andrienko2013visual} builds a framework for spatiotemporal analysis. A rich set of interactive exploration is embedded. Analysts can change the color scheme and assign colors to clusters on maps and line charts. 
Analysts can choose the parameters to be mapped in the parallel coordinates, and adjust smoothing parameters as well as the time period for the contour-based treemap in TelCoVis~\cite{wu2016telcovis}.

Externalization of results records analysts of important discoveries. 
Voila~\cite{cao2018voila} includes a snapshot panel for analysts to conveniently capture the overall and detailed map views. 
Ferreira \etal~\cite{ferreira2013visual} explore taxi trips using TaxiVis, which supports exporting query results in CSV files, the same type of files as their input source. 
The visual analytics framework~\cite{andrienko2013visual} models spatiotemporal data. The model description files can be stored externally along with group membership of place, statistical details.

As analysts gradually develop basic knowledge, they recognize suspicious areas and integrate domain knowledge in anomaly detection. 
After a link is described as anomalous, the link is placed on the top of visualization while the other links become transparent~\cite{ko2014analyzing}. 
In Voila~\cite{cao2018voila}, analysts incorporate their judgments about whether the region is anomalous. This feedback is taken into consideration in the recalculation of anomaly scores of all regions in the space.


\section{Network Communication}
\label{sec:network}
\textit{Network communication} is sending and receiving information between machines via networks.
Examination of network communication has practical significance for national defense~\cite{Foresti2006VisualCO} and commercial enterprises \cite{liao2010visualizing}. Network communication behaviors include routing, network traffic, and port activities, etc. 
Anomaly detection associated with network communication is usually concerned with cyber security, which is protecting computers and systems against malicious activities in a computer-related system. Anomalies are indicated by alarms and suspicious patterns that deviate from expectation. Investigation into these signals reveal attacks such as BGP routing instability~\cite{teoh2004combining, fischer2012vistracer}, virus outbreak~\cite{yin2004visflowconnect}, port scans~\cite{taylor2009flovis, zhao2014mvsec, goodall2005preserving}, and intrusion into systems~\cite{Teoh2004DetectingFA, fischer2014nstreamaware}.

\subsection{Data Types} 
The identified connection between sources and destinations is seen as network data. 
Network data is important for detecting anomalous network communication, as it is the foundation for analyzing information exchange between machines. For example, the network connection between autonomous domains (ASes)~\cite{teoh2002case} and that between subnets and hosts~\cite{lakkaraju2004nvisionip} can be analyzed. 
VisFlowConnect~\cite{yin2004visflowconnect} focuses on network traffic between an internal domain sender and an internal/external domain receiver. 
Liao \etal~\cite{liao2010visualizing} represent enterprise networks consisting of hosts, users, and applications as host-user-application connectivity graphs. From the graphs, the similarity of users by applications can be assessed. 
VisAlert~\cite{livnat2005visualization, Foresti2006VisualCO} considers large-scale attack patterns between alerts and local networks. Analysts can obtain an overview of intrusion attempts and general situations by inspecting networks formed by alerts and a topology map of local network nodes. 

Multidimensional data contains multiple numeric attributes to describe context information in network communication.
Attack frequency, flow rates (\ie number of packets and bytes for a fixed period), and system load are examples of multidimensional data when discussing network communication behavior. 
Teoh \etal~\cite{teoh2004combining} uses intensity, categorical, and counting measures to describe routing behaviors. Each measure has its corresponding degree of abnormality. The anomaly threshold is calculated from the anomaly degrees of multiple measures. SpiralView~\cite{bertini2007spiralview} presents a connection as a list of events introduced in terms of time, source host, application, and destination host. The details of connection are described using multidimensional data, which are incorporated in the description of alarms. 
MVSec~\cite{zhao2014mvsec} uses multidimensional data including the number of connections, flow counts, and flow bytes. The statistics are combined with temporal features to explain each unit of network security data. 

Spatiotemporal data of network communication associates mainly with addresses of receivers and/or senders, and temporal information of occurred activities. 
Spatiotemporal data provides details of timestamps and IP addresses. 
Investigation of spatiotemporal data is helpful for traffic monitoring, as can be seen in~\cite{Mansmann2008VisualAO} which deals with timestamps from millisecond to year together with IP addresses from IP prefix to continents. 
Erbahcer \etal~\cite{erbacher2002intrusion} explore time and difference in IP addresses between the external domain and that of the monitored system. The greater the differences between addresses, the more suspicious the network communication is. 
SpiralView~\cite{bertini2007spiralview} is interested in how alarms evolve in time with the purpose of detecting periodic patterns. By inspecting alarms of the same level of attack severity, alarms can be segmented based on their temporal distribution to better understand network behaviors. 
VisTracer~\cite{fischer2012vistracer} visualizes destination ASes of traceroutes against time to assess spatiotemporal patterns of occurred anomalies.

Text data type provides low-level details about connections in cyber networks. Text data can be encoded to visualization for high-level exploration, or acts as evidence for confirmation of hypothesis regarding anomalousness. Text data includes textual logs and categories of events. 
Erbacher \etal~\cite{erbacher2002intrusion} represent textual log information using glyphs. Textual logs contain time, locations and, types of connection. 
Teoh \etal~\cite{Teoh2004DetectingFA} project connections with known classes (\ie normal, probe, DOS, U2R, and R2L) into regions in a visualization panel. Suspicious data is found separate from normal data, facilitating further investigation.

\begin{figure*}[!t]
    \centering
    \includegraphics[width=\textwidth]{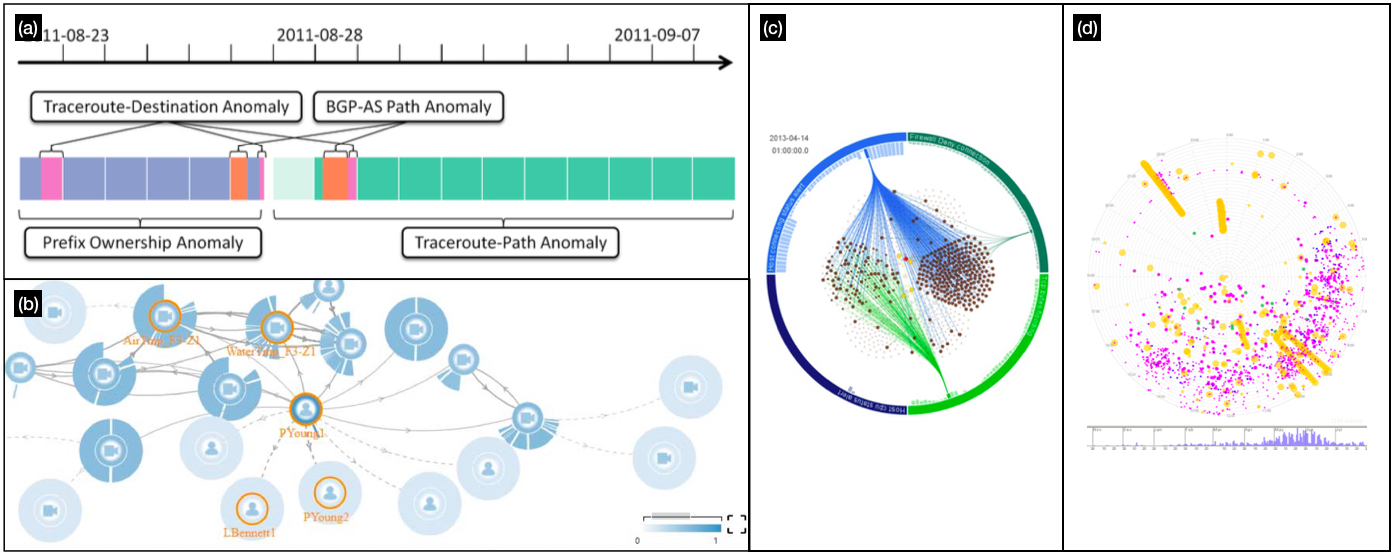}
    \caption{Visualizations of anomalous network communication behaviors. (a) VisTracer~\cite{fischer2012vistracer} visualizes routing anomalies in traceroutes using matrix. (b) Tao~\etal~\cite{tao2018visual} design a high-order correlation graph to show collective anomalies. (c) MVSec~\cite{zhao2014mvsec} mines correlation of events attributed by what, when and where in a dandelion-metaphor using circular-based design. (d) SpiralView \cite{bertini2007spiralview} analyzes how alarms evolve in time and detect suspicious patterns using a radar chart.}
    \label{fig:network} 
\end{figure*}

\subsection{Anomaly Detection Techniques}
Statistical anomaly detection techniques are widely used in the detection of abnormal network communication. Detection techniques of cyber attacks are categorized into signature-based (matching suspicious behaviors with known attack patterns based on existing statistical models or rules) and anomaly-based (comparing behaviors against a ``normal'' baseline)~\cite{Patcha2007AnOO}, both of which can be described using statistical methods. 

We describe visualization works that incorporate statistical methods below. 
Teoh \etal \cite{teoh2004combining} investigate BGP routing instability with a signature-based detection and a statistics-based algorithm. Signatures based on bursts of sequence within a time window are matched with data. The statistics-based approach raises an alarm when current behaviors deviate from expected patterns obtained from history. 
VIAssist~\cite{d2007visual} highlights data instances that meet the criteria of attacks seen in the catalog and discovers the unexpected patterns by interactive exploration of visualization. 
Mansmann \cite{Mansmann2008VisualAO} applies a signature-based algorithm to detecting botnet spread propagation whereas significant traffic changes are visualized in a readily noticeable form. 
VisTracer~\cite{zheng2007light, fischer2012vistracer} compares anomalies with existing scenarios of BGP hijacking. Unknown suspicious attacks are found by adapting online change-point detection algorithm and comparing path similarity. 
MVSec~\cite{zhao2014mvsec} uncovers overall network state details by visualizing several statistical time series including network traffic and the number of distinct active IPs over time. Suspicious patterns are analyzed in terms of what, when, and where from statistics (\eg time interval, flow counts, flow bytes). 
Tao \etal~\cite{tao2018visual} detect point anomalies with a Gaussian model-based technique for labeled data, and with a histogram-based technique for unlabeled data. The correlation analysis and propagation of anomaly score is performed to detect collective anomalies.

Classification-based methods are used in intrusion detection~\cite{Teoh2004DetectingFA, bertini2007spiralview}. 
Teoh \etal~\cite{Teoh2004DetectingFA} utilize a user-directed drawing program, PaintingClass, to classify each object and predict the categories. Unsupervised attacks are found by comparing positions of normal instances and unlabeled data. 
SpiralView~\cite{bertini2007spiralview} models user behaviors using Bayesian networks, and raises anomalies for deviations from usual behaviors.

Nearest neighbor-based techniques based on similarity is applied in~\cite{liao2010visualizing}, which transforms relations among hosts, users, and applications into network connectivity graphs, bipartite graphs, multidimensional scaling, and similarity graphs. The inter-graph similarity is evaluated in a top-down manner, and node similarity is analyzed based on the dynamics of node degrees. 
LongLine~\cite{yoo2018longline} uses local outlier factor to facilitate the comparison of temporal patterns of anomalous systems behaviors. The tool employs a frequency-based model which identifies files and addresses in audit logs as an individual entity. The entity is described by a feature vector constructed from their extended bag of system call models. 

TVi~\cite{boschetti2011tvi} uses a spectral technique to direct users to time periods of anomalous activities. The tool derives a scalable metric (entropy from IP addresses and ports) and conducts dimension reduction using principle component analysis (PCA). 
NStreamAware~\cite{fischer2014nstreamaware} applies a DBSCAN algorithm to cluster timelines, which achieves event detection in streaming data. The possibly important temporal segments are further assessed by analysts through interactive exploration.

\subsection{Visualization Techniques}
\textit{Egocentric Behaviors.} An egocentric network communication behavior triggers alarms due to suspicious network properties of the connection between source host(s) and destination host(s).
Examples of egocentric anomalous network communication behaviors are hijacking network traces by another AS, a port scan, and unusually high volume of traffic on a machine. 
Glyph and graph visualizations are used to represent egocentric behaviors. 

Erbacher \etal~\cite{erbacher2002intrusion} initiated one of the earliest visualizations to display IP addresses of alarms in a glyph-based radial form. Line glyphs surrounding a central node represent different types of connection (\eg parallel lines indicate initial connection requests). The difference in IP addresses between the external domain and that of the monitored system is encoded in the length of line glyphs. The suspicious connection is colored red due to unexpected user activity such as timeouts expire. 
Teoh \etal~\cite{teoh2004combining} inquire into Border Gateway Protocol (BGP) routing instability. Near-real-time monitoring of Internet routing is pictured as temporal line charts and glyphs, where a suspicious event detected from statistics is illustrated with a large circle in high position and a spike in timeline.

Graph visualization, especially matrix is used to detect anomalous egocentric network communication.
Goodall \etal~\cite{goodall2005preserving} develop a matrix showing network activity of hosts over time. Communication between hosts is superimposed on the matrix, complemented by multiple linked views detailing port activity and raw packets. 
NVisionIP~\cite{lakkaraju2004nvisionip} detects traces of abnormal network behaviors in multiple levels of an entire class-B IP network. NVisionIP consists of a galaxy view in matrix, a small multiples view, and a machine view with bar chart. Spikes in traffic volume are seen as changes in node colors in the matrix. Simple scanning attacks are discovered as clusters in the matrix, where x- and y-axis stand for subnets and hosts, respectively. 
VisTracer~\cite{fischer2012vistracer} (Figure~\ref{fig:network}~(a)) tackles large trace route data sets to distinguish legitimate routing changes and spam campaigns. Time and destination of ASes are represented by x- and y-axis in a matrix layout. Rectangular glyphs in the matrix layout are anomalies. Two nearly identical anomaly patterns at the same x-position in the matrix indicate routing anomalies in two ASes. 

\textit{Collective Behaviors.} Collective network communication behaviors involve more than one exchange of information between two machines or among multiple machines. Anomalous behaviors include botnet infection and periodic attacks, which are represented in graph and sequence visualizations.

Tree visualization, one of the graph visualization, helps identify anomalous network communication behaviors. 
Teoh \etal~\cite{teoh2002case} examine routing behavior of BGP data. Each IP address is mapped to one pixel in a quadtree visualization to detect anomalous origin AS changes. An event is represented by a line connecting the affected IP prefix and ASes. Anomalies are revealed as an area concentrated in lines, since events that take similar paths multiple times are suspicious. 
Teoh \etal~\cite{Teoh2004DetectingFA} detect intruders by allowing analysts to interactively explore activity logs in an interactive decision tree visualization layout. Complementary to this view, a three-dimensional scatter diagram pinpoints unlabeled anomalies when a high-density cluster lies in areas of sparse training data. 
Mansmann \etal~\cite{Mansmann2008VisualAO} aggregate IP addresses according to prefix, autonomous system, country and continent in treemaps based on two layout algorithms. This visualization helps monitor large-scale network data. Segments in treemaps are colored indicating sharp changes in the number of incoming connections.

Node-link diagrams visualize structures of collective network communication.
Tao~\etal~\cite{tao2018visual} (Figure~\ref{fig:network}~(b)) design a high-order correlation graph to show collective anomalies. When applied to software analysis, malicious attacks due to software vulnerabilities are identified as collective anomalies. In this case, a node illustrates each line of code, an event represents an execution, and a correlation link represents data flow. 
NIVA~\cite{nyarko2002network, scott2003network} coordinate 3D node-link view with glyph design and circular histograms. It distinguishes from other visualizations as it builds attack severity into interaction inspired by the ``haptic'' concept. For example, when dragging nodes in the three-dimensional view, users can feel the force of ``push'' and ``pull'' motion computed based upon attack frequency. 

Circular-based visualization is also used to demonstrate collective network communication behaviors. 
VisAlert~\cite{livnat2005visualization, Foresti2006VisualCO} identifies critical attacks of hosts through analyzing ``what, when, where'' information of alerts. The alerts are allocated on segments of rings according to the severity of attacks. ``When'' attribute is mapped such that the innermost ring represents the most recent activities. Inside the ring, a network topology map is used to depict network under scrutiny. 
FloVis~\cite{taylor2009flovis} observes interactions between host pairs on either side of the monitored border. A bundle diagram displays connections between entities in a radial tree layout. Scanning activities can be detected by examining bundles directed from 9000 consecutively numbered ports to the internal host. 
MVSec~\cite{zhao2014mvsec} presents four coordinated views to discover anomalies and retrieve stories behind subtle events. The event radar view (Figure~\ref{fig:network}~(c)) mines correlation of events attributed by what, when and where in a dandelion-metaphor in a ring. Seeds (\ie subnets) spread from the center of the dandelion stalk, which represents the only entrance to the network. Antennas (\ie hosts) extend from the seed, giving a two-layer hierarchical structure. The seriousness of botnet infection, for instance, is indicated by the number of colored nodes in the dandelion-metaphor. 

Sequence visualization uncover abnormal trends of collective network communication. 
While NVisionIP~\cite{lakkaraju2004nvisionip} focuses on activities occurred on machines, its complementary tool VisFlowConnect~\cite{yin2004visflowconnect} explores network flows between machines using parallel coordinates. VisFlowConnect investigates the relationship between senders and receivers. A cluster of lines originating from an external host sender indicates a virus outbreak. 
SpiralView \cite{bertini2007spiralview} (Figure~\ref{fig:network}~(d)) analyzes how alarms evolve in time and detect suspicious patterns (\eg alarms appearing everyday at the same time). The alarms are scattered dots in a radar chart, which is useful for identifying periodic patterns of intrusions. The alarms are arranged from the center to the outer part so that recent events are allocated with more space. 
NStreamAware~\cite{fischer2014nstreamaware} analyzes a condensed heterogeneous data stream and uses a sliding slice to provide a summary for the selected period of time. The tool supports omitting and merging normal ranges so that suspicious port activities, attack patterns, and routing behaviors are revealed.

\subsection{Interaction Methods}

Detection of anomalous network communication requires tracking \& monitoring. 
Teoh \etal~\cite{teoh2004combining} direct analysts' attention to anomalies by highlighting the background gray.
In the TVi~\cite{boschetti2011tvi} visual querying system, analysts select an item in the anomaly list, and then the associated time range is highlighted in the timeline visualization. 
In NVisAware~\cite{fischer2014nstreamaware}, analysts can click the star icon to store the real-time sliding slice under investigation. The events marked with star icons are added to the same view. Analysts can determine suspicious patterns from flagged and labeled events from the starred time slices. 
There are four coordinated views in MVSec~\cite{zhao2014mvsec}. Interaction in one view is linked to visualization in another view, which is helpful for digging hidden network attacks that are hard to recognize.

Interesting network communication behaviors are found by exploring visual elements in the same scale or in multiple levels of granularity. 
VisAlert~\cite{livnat2005visualization, Foresti2006VisualCO} enables panning and zooming operations of the topology map in the ring. Analysts can also configure projections onto rings by collapsing and expanding alert grouping on rings. 
Tao \etal~\cite{tao2018visual} employs the direct-walk technique (\ie a series of mouse clicks) for exploring anomalies. When an analyst notices a suspicious node, he/she clicks another node that contributes to the anomaly of the suspicious node. That is, the analyst extends examines effects on the node due to more nodes. 
Mansmann \etal~\cite{mansmann2006interactive, Mansmann2008VisualAO} aggregate IP addresses according to prefix, autonomous system, country and continent in treemaps. Drill-down and roll-up functions can be applied for nodes of the same level of detail.

Interactive methods are used to unveil suspicious patterns of data. 
The filter dialogue in NVisionIP~\cite{lakkaraju2004nvisionip} restricts what data flows to be visualized. Analysts visualize network traffic according to the filters based upon the combination of IP address, ports, protocols, and display type. 
The visual analytics tool FloVis~\cite{taylor2009flovis} has a bundle diagram that describes network flows between a source and a destination. Analysts can loosen the bundles to find suspicious attack patterns. Additionally, analysts can choose to linearly distort points on the circle of the bundle diagram. 
Mansmann \etal~\cite{mansmann2006interactive, Mansmann2008VisualAO} color data in treemaps in a linear or logarithmic scale. Coloring in the logarithmic scale makes the visualization resistant to the randomness of data. 
Teoh \etal~\cite{Teoh2004DetectingFA} use a painting program to help categorize the same type of anomalies into one group. Analysts interactively arrange data instances through drawing, partition, and appropriate coloring.

Analysts may keep a record of results for further analysis. 
The intrusion detection tool NIVA~\cite{scott2003network} allows analysts to export results in an ASCII format.
VIAssist~\cite{d2007visual} is designed for collaborative working environments. The report builder in the visualization tool allows analysts to drag and drop graphical objects in the current display. The results with annotations can then be saved as PowerPoint or PDF file. 
MVSec~\cite{zhao2014mvsec} simplify analysts' operation by offering frequently-used configuration files for anomaly detection. Analysts can export their configurations as a new configuration file.

VIAssist~\cite{d2007visual} has an expression builder and E-Diary to fulfill the refinement \& identification task. Analysts can formulate a hypothesis about a suspicious activity into an expression. A catalog of expressions collects knowledge, \ie hypotheses made by analysts during analysis. The E-Diary helps documentation of hypotheses. This encourages sharing annotations with colleagues and communication of hypotheses in a group. 
Analysts can annotate suspicious patterns in SpiralView~\cite{bertini2007spiralview} for long-term analysis and policy's assessment. The annotations can be an explanation for the anomalies and the action applied to the system.


\section{Transaction}
\label{sec:trasaction}
\textit{Transaction} refers to monetary flows in buying and selling. The goal is to connect financial sources to companies or individuals. 
In a broad sense, stock market deals \cite{huang2009visualization}, credit card transactions \cite{olszewski2014fraud}, business processes \cite{hao2005visimpact, suntinger2008event} are under this category. 
Frauds are the typical type of anomalies associated with transactions, as people may be allured by monetary benefits to perform illegal transactions. Clients may collude with employees in financial institutes in activities of money laundering, unauthorized transactions, and embezzlement, etc. \cite{leite2018eva}. Other anomalies include unexpected business processes \cite{hao2004visbiz, suntinger2008event} and high default group in a network of guaranteed loans~\cite{niu2018visual}.

\subsection{Data Types}
Spatiotemporal data describes details of location, timestamps of transaction, and time series of events. Spatioemporal analysis is critical in financial analysis, and thus detection of anomalous transactions often incorporates analysis of geographic locations and time series. 
Attributes including time of transaction~\cite{suntinger2008event, olszewski2014fraud}, how often a customer executes operations~\cite{leite2016visual} and geographic regions~\cite{hao2006business, leite2018eva} provide a foundation for first-step analysis. 
For example, the Event Tunnel~\cite{suntinger2008event} conducts temporal correlation to link seemingly isolated events, and thus business patterns and fraud patterns involving more than one individual~\cite{huang2009visualization} can be uncovered. 
Huang \etal~\cite{huang2009visualization} perform spatial correlation in addition to temporal and spectral (based on frequency) to identify suspected traders and attack plans. 

Multidimensional data is often used in conjunction with spatiotemporal data to detect anomalous transactions. 
By probing into time series along with details of the amount of money transferred~\cite{leite2018eva, olszewski2014fraud, hao2006business}, the number of transactions within a period of time~\cite{hao2005visimpact, hao2006business}, and number of the activities that are new to the user~\cite{legg2015visualizing}, analysts can gain an overall picture of the histories of financial transactions. 
An example of using multidimensional and spatiotemporal information is VisImpact~\cite{hao2006business}. VisImpact correlates variables of purchase quarter (\ie temporal details), fraud amount, and fraud count to reveal relationships among important factors. 
Legg~\cite{legg2015visualizing} identifies insider threats in an organization by inspecting multidimensional data including the number of times that the user performs particular tasks, number of these activities that are new to this user and to any user in this same position.

Network data describes relationships among entities involved in transactions. A network can be links between traders~\cite{huang2009visualization} in trading networks, between entities such as people, companies, and banks~\cite{didimo2011advanced}, and between enterprises that take loan guarantee~\cite{niu2018visual}.
For example, Niu \etal~\cite{niu2018visual} consider high default groups as communities in networks. A community that interacts with each other internally more frequently than those outside of it can trigger serious financial losses. 
Didimo \etal~\cite{didimo2011advanced} analyze categorical networks that contain different types of entities to discover financial crimes. Indices such as the centrality of a node, like betweenness, and node degree are measured to indicate anomalousness.

When analyzing transaction behavior, categories derived from text help describe the relationship between a payer and a payee~\cite{chang2007wirevis, chang2008scalable}, label different types of activities conducted by employees~\cite{legg2015visualizing}, and identify the type of state changes in a business process~\cite{suntinger2008event}. Text data is used to distinguish between senders, intermediates, and receivers in financial transactions, and to build profiles for analyzing their potential suspicious behaviors. 
For example, WireVis~\cite{chang2007wirevis, chang2008scalable} extracts pre-defined keywords from a set of transactions and relates the keywords that appear in the same transaction. Keyword-to-account relationship is analyzed based on the number of time the keywords appear in that transaction. 
Jigsaw \cite{gorg2013combining} help identify any linkages between people or companies relevant to financial frauds such as fictitious suppliers' invoices and systematic deletion of suppliers' invoices. These linkages are found by keyword/sentence summaries of transactions, sentiment, and word clouds of a document. 

\begin{figure*}[!t]
    \centering
    \includegraphics[width=0.75\textwidth]{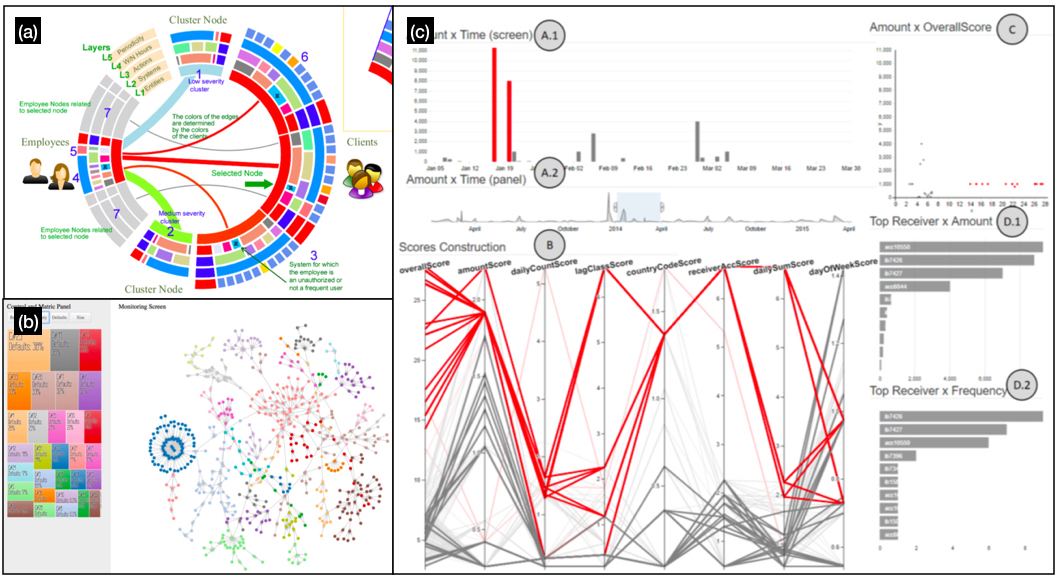}
    \caption{Visualizations of anomalous transaction behaviors. (a) Argyriou \etal~\cite{argyriou2014fraud} use a multi-layer radial drawing to describe activities between employees and clients. (b) Niu \etal~\cite{niu2018visual} assess risk of guaranteed loans by visualizing networks of small and medium enterprises groups using a node-link visualization. (c) Leite \etal~\cite{leite2018eva} design user-friendly views of chart visualizations and parallel coordinates to help identify anomalous transactions.}
    \label{fig:tran} 
\end{figure*}

\subsection{Anomaly Detection Techniques}
Statistical methods applied to the detection of suspicious transactions build normal profiles of customers, and then evaluate new transactions against known anomalies in historical data.
Huang \etal~\cite{huang2009visualization} match suspected patterns in spatial, temporal, and spectral (\ie frequency) domains with similar patterns seen in historical databases, which act as anomalous signatures. 
Leite \etal~\cite{leite2016visual} first build customer profiles from their frequency, amount, and location of transfer from histories. New transactions are then evaluated against the profiles to see if they are anomalous. 
The visualization tool EVA~\cite{leite2018eva} generates customer profiles and provides different statistical measures for new transactions. The statistical profiles combine histograms and rules specified by experts to provide references. Sudden behavior changes in comparison to the profiles are identified as suspicious. Anomalies are highlighted if anomaly scores exceed a threshold.

Application of clustering-based techniques is based on the assumption that anomalous financial communities share common features within a group. 
WireVis~\cite{chang2007wirevis, chang2008scalable} implements the k-d tree algorithm to detect suspicious behaviors in wire transactions. It treats accounts as points in k-dimensional space, where k is the number of attributes. The accounts are grouped using a centroid-based clustering technique. 
Schaefer \etal~\cite{schaefer2011visual} cluster entries based on similarity of temporal event patterns so that analysts can identify suspicious patterns in a packed visualization. An event pattern refers to an event sequence or event episode that displays interesting properties. 
Didimo \etal~\cite{didimo2011advanced} apply hierarchical clustering by finding k-cores in a graph, which is effective for discovering relevant groups in networks. This graph-based clustering defines clusters of cohesive structures, in which each cluster has at least k inter-connected neighboring points. 
Clustering based on graph structure is used in Network Explorer~\cite{guerra2016network}. Communities in the financial network can be identified as clusters converted from undifferentiated nodes and edges. Two clustering algorithms are employed to process large-scale networks on the server side and process smaller networks on the client side.

Classification-based, nearest neighbor-based, information theoretic, and spectral techniques are discussed below. 
Olszewski \cite{olszewski2014fraud} uses a threshold-type binary classification technique to determine whether an account in self-organizing maps (SOM) is fraudulent or not. The threshold is computed by measuring dissimilarity between the centroid of SOM grid and the maximal value in the matrix. Accounts with values higher than the threshold are anomalous. 
A decision tree \cite{argyriou2014fraud} is generated from the patterns suggested by auditors. To detect internal frauds conducted by employees, each employee is assigned an anomaly value. The value indicating the severity of anomalousness is obtained by evaluating event series of an employee against fraud patterns. 
Structured networks reveal anomalies. Two risk indices~\cite{didimo2018visual} based on neighborhood structure, \ie pattern centrality and transaction pattern centrality, are computed by assigning weights to each edge that corresponds to a taxpayer in a transaction network.
Niu \etal~\cite{niu2018visual} employ an information theoretic-based approach to uncover risk guarantee pattern and detect high default groups for loans risk management. Specifically, the proxy for information flow is the probability flow of random walks in directed weighted networks. 
PCA is utilized for identifying insider threat~\cite{legg2015visualizing} due to its effectiveness in detecting users that exhibit irregular variances across the set of derived features. An interactive PCA helps comprehend relationships between the PCA space and the original higher-dimensional space in a visual interface.

\subsection{Visualization Techniques}
\textit{Egocentric Behaviors.} An egocentric transaction is described as buying or selling behaviors conducted by an individual. An anomalous egocentric transaction can be an unauthorized transaction or a deal with an exceptionally high amount of value. Detection of these behaviors mainly uses sequence visualization.

VisImpact~\cite{hao2005visimpact, hao2006business} organizes attributes of transactions by allocating them onto three parts/axes of a ring: left semicircle, bisector, and right semicircle. Each axis stands for an attribute of interest (\eg region, client, fraud amount, fraud count). 
Suntinger \etal~\cite{suntinger2008event} display events as nodes in a cylindrical tunnel. The top view of the cylinder represents historical events, which are laid out such that more recent events are in the outer ring. Details of events are encoded by the color and size of glyphs of the Event Tunnel. Anomalous betting behaviors of a user are discovered by temporally correlating the account history events of the user to known suspicious account profiles. 
Argyriou \etal~\cite{argyriou2013occupational} study the temporal relationship of transactions between a pair of client and employee in a radar chart. The nodes in the radar chart represent transactions, which are positioned according to the time of action, pre-defined periodicity, and ordering of timelines. Events/transactions related to the same client along the radius of the radar chart are considered suspicious, as the patterns suggest the employee falsifies the client's invoices.  

Graph and text visualizations are also used to demonstrate suspicious egocentric transaction behaviors. 
Argyriou \etal~\cite{argyriou2014fraud} (Figure~\ref{fig:tran}~(a)) use a multi-layer radial drawing to describe activities between employees and clients. Each layer represents a pattern that is suspicious in different aspects (\eg actions, systems, periodicity), with heat maps in the side view measure anomalousness. When an employee is found to perform events that share similarity with fraud patterns, a suspicious egocentric behavior is identified. 
Jigsaw~\cite{gorg2013combining} mines relationships between entities in text documents. The parallel coordinates view reveals the correlation of selected attributes (\eg company, person). By combining with the heat map for sentiment/similarity analysis, cluster view for grouping similar documents, and document view for details, anomalous behaviors can be detected from unique text entities. 
Following that work, Kang \etal~\cite{kang2012examining} studies applications of Jigsaw in various situations including financial transaction. An employee's egocentric behavior of creating fictitious supplier invoices was discovered.

\textit{Collective Behaviors.} A collective transaction behavior involves several parties in transaction and businesses. Collective transaction behaviors include a series of wire transfer and periodic transaction. Graph visualizations are popular among research works interested in transaction behaviors. 

Graph visualization is popular for uncovering collective transaction anomalies. 
Huang \etal~\cite{huang2009visualization} develop two stages to inspect stock market security. Firstly, market performance is evaluated using three-dimensional treemaps, with the heights of blocks indicating the current price of stocks. Secondly, trading networks are compared against suspicious patterns in the historical database. Structured networks are regarded as collective anomalies in transactions. 
Several visual analytics tools~\cite{didimo2011advanced, didimo2018visual, niu2018visual, guerra2016network} develop categoric node-link visualizations where analysts can merge, split, define new subgraph structure, cluster nodes by top-down or bottom-up paradigm, and adjust node sizes by a chosen measurement. Users edit networks interactively to discover communities, which are signals for suspicious financial transactions. 
Didimo \etal~\cite{didimo2011advanced} detect financial activity networks such as money laundering by illustrating entities involved in transactions with nodes. The entities include banks, companies, persons, bank accounts, transactions, and reports filing. Edges between nodes represent semantic connections. For instance, two disjoint clusters that indicate fraudulent patterns are revealed after clustering. The level of depth of a cluster reflects the extent of criticism of the illegal activity. 
Niu \etal~\cite{niu2018visual} (Figure~\ref{fig:tran}~(b)) assess the risk of guaranteed loans by visualizing networks of small and medium enterprises groups which back each other to enhance the financial security. Anomalies, \ie high default groups, are identified as communities in the network using a node-link visualization. A complementary treemap supports navigation of labels/categories and presentation of default rates. 

Chart and sequence visualizations are also used to detect collective transaction behaviors. 
WireVis~\cite{chang2007wirevis, chang2008scalable} uses multiple coordinated chart visualization to analyze suspicious wire transfers between a payer to a payee via a chain of intermediaries. The overall trends of activities and individual transactions are represented by strings and beads in an x-y plot of transaction value against time. Suspicious transactions are the ones relevant to a keyword that is only found in the second half of the year, and a transaction of much higher value than others. 
Leite \etal~\cite{leite2018eva} (Figure~\ref{fig:tran}~(c)) design user-friendly views of chart visualizations and parallel coordinates to help identify the anomalous connection between the amount and the suspicious transactions. If anomaly scores of transactions deviate from normal ranges, the days that contain at least one suspicious transaction are highlighted in red.

\subsection{Interaction Methods}
Analysts track suspicious data by highlighting and correlating relevant data.  
The visual analytics tool EVA~\cite{leite2018eva} computes the overall anomaly scores and sub-scores according to different standards. If the overall score of transactions exceeds a threshold, the transactions are highlighted in red in the parallel coordinates view. Also, selection in another coordinated chart highlights associated transactions and gray out others in the parallel coordinates view. 
When analysts click a node of interest, relevant data that are originally not visualized is displayed~\cite{didimo2011advanced}. This helps analysts discover interesting features that are not apparent from one view, and identify different relationships between data instances. 
A similar operation is seen in~\cite{argyriou2014fraud}, where the selection of one node adds related employees (\ie nodes) into the visualization. Thus, frauds carried out by two or more employees can be tracked.

VisImpact~\cite{hao2005visimpact, hao2006business} supports simultaneous browsing and navigation of multiple nodes. Details of a single node representing a transaction record can be obtained using the drill-down function. 
For the transaction of an account, transactions can be aggregated in terms of day, week, or month in WireVis~\cite{chang2007wirevis, chang2008scalable}. Zooming is enabled in the heat map and temporal chart view. One can also drill down to individuals and compare their records against each other in WireVis. 
Network Explorer~\cite{guerra2016network} includes an overview and an egocentric mode which detects important clusters and individual nodes, respectively. In the overview mode, analysts can navigate to one cluster and compute sub-communities on demand. In the egocentric mode, analysts navigate nodes using the direct-walk from a starting point.

Pattern discovery is often used to help identify anomalous behaviors. 
Filtering in WireVis~\cite{chang2007wirevis,chang2008scalable} is conducted using a set of keywords and criteria like amounts of words. Analysts can select reasonably sized subsets for re-clustering to generate clusters that exhibit interesting features. Furthermore, the color scheme is chosen depending on the characteristic (\eg sequential or diverging) of the measurement in the heat map. 
Jigsaw~\cite{gorg2013combining} allows involvement in defining clusters of text documents, removing false positives, adjusting the number of words shown, and reordering the entity list. 
Dragging, merging, and splitting visual elements are often seen in node-link visualization~\cite{didimo2011advanced, didimo2018visual, niu2018visual, guerra2016network}. 
To discover the tax evasion behaviors~\cite{didimo2018visual}, analysts can merge and split node-link representation. A selection of subgraphs is ranked according to criteria such as the total amount of economic transactions or the risk index. Additionally, analysts can define and draw suspicious graph patterns using pre-defined operators.

A few visualization tools support exporting analyzed results. 
The Event Tunnel~\cite{suntinger2008event} contains a snapshot management console that captures the current state and configuration.
Argyriou \etal~\cite{argyriou2013occupational, argyriou2014fraud} design the exporting function in the visual analytics tools for detecting occupational frauds. The ranking results of anomalousness can be exported in separate log files. The visualization containing suspicious transaction patterns can be stored for post-analysis.

Visual analytics involve domain knowledge into the process of anomaly detection.
Analysts are enabled to reassign labels of the ``structure hole spanner'' during interactive exploration~\cite{niu2018visual}. The structure hole spanner interlinks different communities in a network, which can be modified through merging and splitting operations. High default groups are found to be associated with these labels. 
In TAXNET~\cite{didimo2018visual}, analysts can define graph patterns based on their understanding of tax evasion frauds. Textual labels are attached to the graphs to describe rules for nodes (\ie taxpayers) or edges (\ie relationship).

\section{Discussion and Outlook}
\label{sec:discussion}
In this section, we first summarize trends of research interest in the community of data visualization regarding anomalous user behaviors. We then discuss our findings regarding data types, anomaly detection techniques, visualization techniques, and interaction methods across different user behaviors.

\subsection{Visual Analytics of Anomalous User Behaviors}
Visual analytics of private \textit{social interaction} behaviors related to emailing received substantial attention in 2000s but showed significant decreases since then. Recent research works~\cite{redondo2015layer, van2014dynamic} are more interested in the social network structures found in emailing, calling behaviors.  
A clear trend worth noticing is the popularity in analyzing public social interaction behaviors related to posting in social media since 2010. The volume of social media data ensures wide coverage of people's behaviors including anomalous and normal behaviors. Application to real world is attractive from the perspective of social science and possibly more. 
We have seen many visualization tools that address event detection from massive information, information spreading, and identification of social bots. However, to the best of our knowledge, we found that only a few visualization works~\cite{sun2018fraudvis} focuses on secretive or collusive anomalous behaviors, when compared to machine learning approaches~\cite{jiang2016suspicious} that detect suspicious behaviors. Specifically, we have not seen visual analytics methods for detecting social Sybil attacks (\ie astroturfing)~\cite{yu2008sybillimit} or private information inference~\cite{heatherly2013preventing} related to the posting behavior. We are hoping to see more efforts to be put in discovering anomalous behaviors conducted in a collusive, secretive manner.

As for \textit{network communication}, the research interest remains relatively strong, though classical works~\cite{yin2004visflowconnect, lakkaraju2004nvisionip} that analyze this behavior are mostly published in 2000s. 
Visual analytics of network communication focuses on aggregating different levels of data as well as real-time monitoring. 
Aggregation of data is often used to monitor high-level structures of networks and at the same time, to visualize anomalies in an interface of limited space. As data sources of audit logs and network traffic provide detailed and systematic information, attacks are often traceable to individual machines even though malicious activities originate from more than one device. 
In addition, the preference for real-time or near-real-time monitoring in intrusion detection~\cite{Mukherjee1994NetworkID} is emphasized, manifested by the realization of analyzing streaming data in many visualizations. This results from the need for timely detection of malicious attacks. As computing abilities advance, we expect to see more visualization tools that can handle streaming data.

\textit{Travel} receives continuous attention of researchers given that more data is available for analysis (mobile phones~\cite{von2012visual}, geo-located messages~\cite{thomas2009challenges}, maritime search and rescue events~\cite{riveiro2009interactive}).
Though visualization techniques used for analyzing travel behaviors are similar (\ie geographic visualization), a rich set of interaction methods is implemented in order to detect and comprehend anomalies~\cite{von2012visual, ferreira2013visual}. By analyzing patterns in user-specified spatial and temporal ranges, analysts study user behaviors in multiple levels of granularity to and fro, and gradually develop their understanding during interactive exploration. As more and more sensors are available in daily life, we hope to finer segmentation of groups of people to offer an accurate description of travel patterns.

Visualization works regarding anomalous \textit{transaction} behaviors modernizes traditional visual methods in the financial field. For example, EVA~\cite{leite2018eva} integrates human decisions into the analysis of frauds into the existing alert system.
In recent years, we have seen an increased number of visualization tools designed for detecting suspicious users involved in financial transactions. However, by comparing the average number of citations between user behaviors, the overall research interest in financial transactions is less than those in travel behaviors, for example. Privacy issues can largely limit the resources available for research. Having said that, we are hoping to see more in-depth collaboration between academic researchers and financial institutes to resolve transaction frauds by recognizing fraudsters' behaviors. 

\subsection{Data Types}
Application of \textit{multidimensional data} to anomaly detection can be found across four behaviors. It offers a variety of features for detecting anomalous behaviors and is often used in conjunction with other data types.
\textit{Text} is an important data type for detecting abnormal social interaction behaviors, whereas text is a compliment in the analysis of other user behaviors. Text provides information about identities and backgrounds of objects involved, which is used to categorize objects. 
\textit{Network} is used frequently in the analysis of network communication as well as social interaction behaviors. Links exist in cyber networks between sources and destinations, and social networks between senders and receivers.
\textit{Spatiotemporal information} enriches skeletons of analysis by incorporating contextual information of users' travel behaviors. Detection of anomalous transaction and social interaction behaviors often incorporates temporal analysis. 

Analysis based on data types helps indicate overlapping areas between user behaviors, which is a signal of borrowing analytics approaches from other behaviors. For example, exploration of rating behaviors in online e-commerce stores is similar to that of network security problems. Sensitivity to time-critical behaviors in anomaly detection is emphasized in~\cite{webga2015discovery}, in which streaming data is processed. Network between sources and destinations is found in network communication, whilst network between users and items is also important for discovering rating frauds. 
We see a trend of incorporating multiple types of data. Since anomaly detection problems often encounter unknown ill-defined anomalies, usage of all four data types can create a relatively thorough picture for investigation.

\subsection{Anomaly Detection Techniques}
\textit{Statistical} techniques are most widely used. The principle of employing statistical techniques is more intuitive compared to the other techniques: data that are not described by the known distribution are anomalous. 
For example, a majority of network communication behaviors are studied using statistical techniques. Detection techniques for cyber attacks are classified into signature-based and anomaly-based~\cite{Patcha2007AnOO}, both of which can be applied with statistical-based approaches.
\textit{Clustering-based} techniques are often used in studying travel and transaction behaviors. Clustering is often employed to tackle large-scale databases associated with travel behaviors. Clustering methods in transactions divide customers into groups based on the assumption that abnormal behaviors are found outside the clusters. 
\textit{Nearest neighbor-based} techniques are applied to detecting anomalous social interaction behavior. For example, in graphs composed of senders and/or receivers in associated with emailing and calling, anomaly scores are computed from distance or densities. 


We expect to see more visualization tools to employ anomaly detection techniques such as machine learning approaches in the future. The effectiveness of machine learning methods in visualization is well-recognized \cite{liu2017towards}. Though the time interval between the release of detection techniques and the implementation in visualization might be long (\eg a five-year interval for FraudVis \cite{sun2018fraudvis} to apply the CopyCatch \cite{beutel2013copycatch} algorithm), we believe machine learning techniques are of great value for anomaly detection in visualization. 
Recently, Chalapathy and Chawla survey \cite{chalapathy2019deep} deep learning techniques for anomaly detection. 
For example, Malhotra \etal ~\cite{malhotra2016lstm} develop a Long Short Term Memory Networks based Encoder-Decoder scheme for Anomaly Detection (EncDec-AD) that is able to uncover predictable, unpredictable, periodic, and aperiodic in long and short time series. 
Anomalies in multivariate time-series data are uncovered using a Multi-Scale Convolutional Recurrent Encoder-Decoder (MSCRED)~\cite{zhang2018deep}, which can capture dynamics and encode the inter-correlations between different pairs of time series.

\subsection{Visualization Techniques}
Among \textit{graph visualization}, node-link diagram is mostly used in social interaction, transaction, and network communication. Node-link diagram is advantageous in its traceability from one node to the other. It is capable of tracking down to abnormal individuals from email and call records, to individual machines in malicious cyber attacks, and to a pair of employee and client in financial frauds. 
\textit{Text visualization} is favored in the analysis of public social interaction behaviors such as posting. These visualization tools are usually equipped with views showing text data to enable interactive exploration and affirmation of suspicious events or users. For example, to complement inspection of microblogs, original messages and keywords are often found in a table format or tag clouds~\cite{sun2013visitpedia, cao2016targetvue}.
Detection of anomalous transaction behaviors also uses \textit{sequence visualization} such as parallel coordinates. Variations of the relationship between subsequent events can be tracked by changes of linkage between two successive axes, which suggest suspicious transactions occurred. Varied configurations of parallel coordinates include radar chart and Sankey diagram. To illustrate social interaction behaviors, changes of heights and size of bubbles in timeline visualization are used to encode sudden and/or important changes in the volume of keywords. 
\textit{Geographic visualization} is often used to represent travel behaviors as it has the advantage of illustrating two-dimensional physical movement. Flows and bubbles projection on a map show differences in traveling directions and spatial densities of distribution. Heat map is popular to demonstrate spatial densities of humans and vehicles, as it minimizes visual occlusion that may happen in flows/bubbles projection on maps. 
\textit{Chart visualization} is effective in illustrating well-understood anomalies as long as dimensions of the displays are selected properly.

We also found that the number of visualization works that address egocentric behavior forms is much fewer than those studying collective behavior forms. 
Glyph visualization is suited to visualizing egocentric behaviors as differences in individuals' roles can be identified more efficiently. Visualizations of collective behaviors take a variety of representations
To better explain, we use an example in social media where the same user behavior results in problems viewed from egocentric and collective perspectives, respectively. 
Both FluxFlow~\cite{zhao2014fluxflow} and Episogram~\cite{cao2016episogram} analyze retweeting behaviors in Twitter. FluxFlow emphasizes the information diffusion process and visualizes temporal evolution of a group of retweeted microblogs using packed colored circles. Episogram, on the other hand, considers whether a Twitter account is anomalous by comparing one's individual retweeting patterns with others'. A user is represented as a glyph, which is later found to be used as a typical visualization for egocentric behavior form.

The trend of applying visualization techniques to detecting anomalous user behaviors is summarized as follows.
Node-link diagram has long been a popular choice of visualizing anomalous user behaviors. It is still a favored technique as it is effective to present an overall structure as well as detailed information when incorporated with rich interaction techniques. Circular-based designs are gaining attention from researchers for its ability to show connections in a packed visualization, where hierarchical structure is displayed using bundles and tree layout inside the ring. Also, circular-based designs usually represent structures of larger-scale than those (\eg stars, cliques) in node-link structures.

We observed an increasing trend of using heat maps when compared to flows/bubbles/3D projection on a map. The reason may be that flows/bubbles/3D map result in visual occlusion, which can only be resolved with appropriate interaction techniques. The opposite trend to that of heat map can be explained by its potential to visualize large-scale data with geographic references. It is able to encode some degree of geographical information, and at the same time, variables such as density of users, anomaly degree can be encoded on the map without occlusion. 
Interest in applying chart visualization has decreased in recent years. Chart visualization is restricted to a few variables, which is ineffective in anomaly detection when an analysis of multiple variables is required.

\subsection{Interaction Methods}
\textit{Exploration \& navigation} has been the most popular interaction task in visual analytics of anomalous user behaviors. 
Most visualization tools support users to gain a high-level summary of large-scale data first and then drill down to anomalies on request. 
The second most popular interaction task is \textit{tracking \& monitoring}. As the papers surveyed are related to anomaly detection, keeping track of suspicious spots is important during interactive exploration. Analysts also highlight data of interest to show its correlation between in the coordinated views, which helps form a picture of where anomalies originate from. 
\textit{Pattern discovery} is also frequently used. 
During the process, the visual representation of data changes accordingly. These updates of one's knowledge drive analysts to construct hypotheses of anomalies.

We observe trends of utilizing interaction tasks in different user behaviors. 
Visualization works that study travel behavior often incorporate exploration \& navigation in map visualization. The reason is that panning on a map is seen often when tracking physical movement~\cite{cao2018voila, andrienko2013visual}. 
Pattern discovery illustrates more than one abnormal feature of anomalies by changing color spectrum and representing traveling patterns in various forms on a map~\cite{chae2014public, ferreira2013visual}. Also, filtering by keywords is seen in social interaction~\cite{webga2015discovery, cao2016targetvue, perer2006balancing, joorabchi2010emailtime} where textual contents are important for determining anomalies. 
\textit{Knowledge externalization} is usually seen in network communication~\cite{d2007visual, zhao2014mvsec} and transactions \cite{chang2007wirevis, argyriou2014fraud}. 
This interaction task enables the processed results to be outputted for further analysis and validation with domain experts.

We increasingly see visualization tools involve \textit{refinement \& identification} in rendering visualization. 
This type of interaction goes beyond the definition of interaction methods~\cite{yi2007toward} because adjustments in anomaly detection algorithms are allowed (\eg Filter technique). 
Several research works allow analysts to adjust parameters in constructing queries~\cite{ferreira2013visual, bosch2013scatterblogs2}, changing thresholds of anomalies~\cite{webga2015discovery, von2012visual}, and updating feedback in anomalies~\cite{cao2018voila}. Visual representation is modified due to fundamental calculation rather than the adjustment of visual encoding. These works facilitate visual analytics by involving human perception and interpretation into the computation process of anomaly detection, which is a deeper level of computer-human interaction than those identified in ~\cite{yi2007toward}.

\section{Conclusion}
\label{sec:conclusion}
In this work, we present a survey of visual analytics of anomalous user behaviors. 
We analyze the related the-state-of-art according to the proposed taxonomies.
Our survey suggests trends and preferences in data types, anomaly detection techniques, visualization techniques, and interaction methods. With these findings, we also highlight potential research directions. 
We believe our work shed light on understanding and analyzing anomalous user behaviors using visual analytics approaches.

\section{Acknowledgments}
Nan Cao is the corresponding author. This research was sponsored in part by the Fundamental Research Funds for the Central Universities in China.


\end{document}